\documentclass{aa}
\usepackage[varg]{txfonts}
\usepackage{graphicx}
\usepackage{gensymb}
\usepackage{appendix}
\usepackage{bigints}
\usepackage{placeins}
\usepackage{soul}
\usepackage{hyperref,natbib,txfonts,xcolor}
\hypersetup{
    colorlinks,
    citecolor=blue,
    linkcolor=blue,
    urlcolor=blue
}

\begin{document}

\title{Magnetic reconnection plasmoid model for Sagittarius A* flares}

\author{N. Aimar\inst{1} \and A. Dmytriiev\inst{2} \and F. H. Vincent\inst{1} \and I. El Mellah\inst{3} \and T. Paumard\inst{1} \and G. Perrin\inst{1} \and A. Zech\inst{4}}
\institute{LESIA, Observatoire de Paris, Université PSL, CNRS, Sorbonne Université, Université Paris Cité, 5 place Jules Janssen, 92195 Meudon, France\\
              \email{nicolas.aimar@obspm.fr}
         \and
           Centre for Space Research, North-West University, Potchefstroom, 2531, South Africa
         \and
            Univ. Grenoble Alpes, CNRS, IPAG, 38000 Grenoble, France
         \and
             LUTH, Observatoire de Paris, CNRS, Université Paris Diderot, 5 place Jules Janssen, 92190 Meudon, France\\
             }

\date{Received September 09, 2022; accepted January 27, 2023}

\abstract{Sagittarius A*, the supermassive black hole at the center of our galaxy, exhibits episodic near-infrared flares. The recent monitoring of three such events by the GRAVITY instrument has shown that some flares are associated with orbital motions in the close environment of the black hole. The GRAVITY data analysis points at super-Keplerian azimuthal velocity, while (sub-)Keplerian velocity is expected for the hot flow surrounding the black hole.
}
{
We develop a semi-analytic model of Sagittarius~A* flares based on an ejected large plasmoid, inspired by recent particle-in-cell global simulations of black hole magnetospheres. We model the infrared astrometric and photometric signatures associated to this model.
}
{
We consider a spherical macroscopic hot plasma region, that we call a large plasmoid. This structure is ejected along a conical orbit in the vicinity of the black hole. This plasmoid is assumed to be formed by successive mergers of smaller plasmoids produced through magnetic reconnection that we do not model. Non-thermal electrons are injected in the plasmoid. We compute the evolution of the electron-distribution function under the influence of synchrotron cooling. We solve the radiative transfer problem associated to this scenario and transport the radiation along null geodesics of the Schwarzschild spacetime. We also take into account the quiescent radiation of the accretion flow, on top of which the flare evolves.
}
{
For the first time, we successfully account for the astrometric and flux variations of the GRAVITY data with a flare model that incorporates an explicit modeling of the emission mechanism. We find good agreement between the prediction of our model and the recent data. In particular, the azimuthal velocity of the plasmoid is set by the magnetic field line it belongs to, which is anchored in the inner parts of the accretion flow, hence the super-Keplerian motion. The astrometric track is also shifted with respect to the center of mass due to the quiescent radiation, in agreement with the difference measured with the GRAVITY data.
}
{These results support the picture of magnetic reconnection in a black hole magnetosphere as a viable model for Sagittarius~A* infrared flares.}

\keywords{Accretion, accretion disk - Magnetic reconnection - Black hole physics - Relativistic processes - Radiative transfer - Radiation mechanisms: non-thermal}
\maketitle

%

\section{Introduction}
The Galactic Center hosts the compact radio source Sagittarius A* (Sgr A*) with an estimated mass of 4.297 million solar masses at a distance of only 8.277 kpc \citep{2022A&A...657L..12G}. This makes the compact object associated to Sgr A* the closest supermassive black hole (SMBH) candidate to Earth. Sgr~A* is a low-luminosity accretion flow with an accretion rate of $(5.2-9.5) \times 10^{-9} M_\odot \ yr^{-1}$ and a bolometric luminosity of $(6.8-9.2) \times 10^{35}$ erg s$^{-1}$ \citep{2019ApJ...881L...2B, 2022ApJ...930L..16E} and thus is accreting at a very sub-Eddington rate. It has been the subject of numerous observing campaigns over the past two decades in order to test the massive black hole (MBH) paradigm (see \citet{2020A&A...636L...5G}) and study the physics of radiatively inefficient accretion flows (RIAF) around SMBH.

Sgr A* shows a slow and low amplitude variability in radio \citep{1975ApJ...202L..63L, 1978ApJ...222L...9B, 1998A&A...335L.106K, 1999ASPC..186..113F, 2006ApJ...648L.127B, 2021MNRAS.505.3616M}, in millimetre and submillimetre \citep{2005ApJ...623L..25M, 2006ApJ...646L.111M, 2006ApJ...650..189Y, 2008ApJ...682..373M, 2015A&A...576A..41B, 2022ApJ...930L..19W}, but also large amplitude and rapid variability in near infrared \citep[NIR;][]{2003Natur.425..934G, 2004ApJ...601L.159G, 2007ApJ...667..900H, 2014ApJ...793..120H} and in X-rays \citep{2001Natur.413...45B, 2012ApJ...759...95N, 2013ApJ...774...42N, 2014ApJ...786...46B, 2015MNRAS.454.1525P}. The flux distribution in the NIR of Sgr A* has been the subject of numerous studies. Some claim a single state modeled by rednoise \citep{2018ApJ...863...15W, 2019ApJ...882L..27D} for the variability of Sgr A* while others claim that there are two states for Sgr A* \citep{2003Natur.425..934G,2011ApJ...728...37D,2020A&A...638A...2G, 2021ApJ...917...73W}: a continuously low amplitude variable state called "quiescent state" and the "flare state" described by short and bright flux with a typical timescale of 30 minutes to 1 hour with a rate of $\sim$ 4 a day. Multi-wavelength studies show that when an X-ray flare is observed, there is a counterpart in NIR suggesting a common origin but the reverse is not true \citep{2018ApJ...864...58F}. Moreover, the flare can also be observed in sub-mm but with a time lag of several minutes \citep{2008A&A...492..337E, 2009A&A...500..935E, 2009ApJ...698..676D, 2021ApJ...923...54M, 2021ApJ...917...73W} following a dimming \citep{2022ApJ...930L..19W, Ripperda21}.

Recently, the GRAVITY instrument \citep{2017A&A...602A..94G, 2008poii.conf..431E, 2011Msngr.143...16E, 2008poii.conf..313P} was able to resolve the motion of the NIR centroid during three bright flare events, showing a clockwise, continuous rotation at low inclination close to face-on ($i \sim 20 \degree$) consistent with a region of emission located at a few gravitational radii $r_g = GM/c^2$ from the central black hole \citep{Gravity2018}. These flares are thus powered very close to the event horizon of the black hole. The exploration of a relativistic accretion region as close to the event horizon with high-precision astrometry and imaging techniques like GRAVITY and the Event Horizon Telescope (EHT) \citep{2022ApJ...930L..12E} promises important information for physics and astronomy, including new tests of the MBH paradigm.

Significant efforts have been made to explain the flares of Sgr A*: rednoise \citep{2009ApJ...691.1021D}, hot spot \citep{2009ApJ...692..902H,2003Natur.425..934G,2006MNRAS.367..905B}, ejected blob \citep{2014MNRAS.441.3477V}, star-disk interaction \citep{2004A&A...413..173N} and disk instability \citep{2006ApJ...636L..33T}. 
The GRAVITY observations in 2018 \citep{Gravity2018} support the hot spot model. However, the physical origin
of such hot spots remains an open question.
Instabilities in black hole accretion disks are a candidate,
for instance the triggering of
Rossby Waves Instabilities \citep[RWI;][]{2006ApJ...636L..33T, 2014MNRAS.441.3477V}.
Alternatively, it could originate from the dissipation of electromagnetic energy through magnetic reconnection. This modification of the magnetic field topology results from the inversion of the magnetic field orientation across a current sheet which eventually breaks into magnetic islands called plasmoids \citep{Komissarov04, Komissarov05, Komissarov07, Loureiro07, Sironi14, Parfrey19, Ripperda20, 2021MNRAS.502.2023P}. In the past years, numerical simulations have repeatedly highlighted the ubiquity of magnetic reconnection in BH magnetospheres, whatever the physical point of view adopted: global particle-in-cell (PIC) simulations in Kerr metrics \citep{Cerutti, 2022PhRvL.129t5101C}, resistive general-relativistic magneto-hydrodynamics (GRMHD) simulations \citep{Ripperda20, Dexter20b, Dexter20a} or resistive force-free simulations \citep{Parfrey+15}. PIC simulations show that magnetic reconnection in the collisionless corona of spinning BHs can accelerate leptons up to relativistic Lorentz factors of $\gamma\sim10^{3...7}$ \citep{Cerutti}, sufficiently high to generate the variable IR (and X-ray) emission \citep{2017ApJ...850...29R, Werner18, 2018ApJ...862...80B, 2021ApJ...922..261Z,2022MNRAS.511.3536S}.

The GRMHD and PIC frameworks each have different limitations. GRMHD simulations describe the evolution of the accretion flow over long time scales, typically of the order of several 100,000 $r_g/c$, but they rely on a fluid representation. Consequently, they cannot self-consistently capture the kinetic effects which are important to constrain dissipation, particle acceleration and subsequent non-thermal radiation. On the other hand, PIC simulations provide an accurate description of the microphysics but at the cost of simulations which can only span a few 100$r_g/c$ in time and with limited scale separation between global scales and plasma scales.

We develop a semi-analytical
model, fed by the knowledge accumulated by recent GRMHD and GRPIC simulations.
The aim is to condense into a reasonably small set of simple parameters
the complex physics of GRMHD and GRPIC models, and thus allow to probe
a large parameter space within a reasonable computing time. 
We also want to remain as agnostic as possible 
regarding the initial conditions of the flow.
In this context,
we discuss the interpretation of the~\citet{Gravity2018}
flare data paying particular attention to the following diagnostics:

\begin{itemize}
\item the marginally detected shift between the astrometric
  data and the center-of-mass location;
\item the tension between the data and the hot spot
  model used by~\citet{Gravity2018}, which assumes a Keplerian orbit;
\item the physical origin of the rising and decaying
  phases of the flare light curve in the context of magnetic reconnection.
\end{itemize}

The first point can be discussed in the context of a very simple
hot-spot model and is the main topic of Sect.~\ref{sec:hot spot+ jet}.
Sect.~\ref{sec:Plasmoid} is the core of our study and focuses on
the second and third points above. It presents
a semi-analytical large plasmoid model due to magnetic reconnection.
It highlights in particular the impact of considering a self-consistent
evolution of the electron distribution function through  kinetic modeling.
This section shows that our plasmoid model is able to reasonably
account for the \citet{Gravity2018} flare data.
The limitations of our plasmoid model are discussed in
Sect.~\ref{sec:limitation}.
The conclusions and perspectives are given in Sect.~\ref{sec:discussion}.

\begin{figure*}[ht]
    \centering
    \includegraphics[scale=0.8]{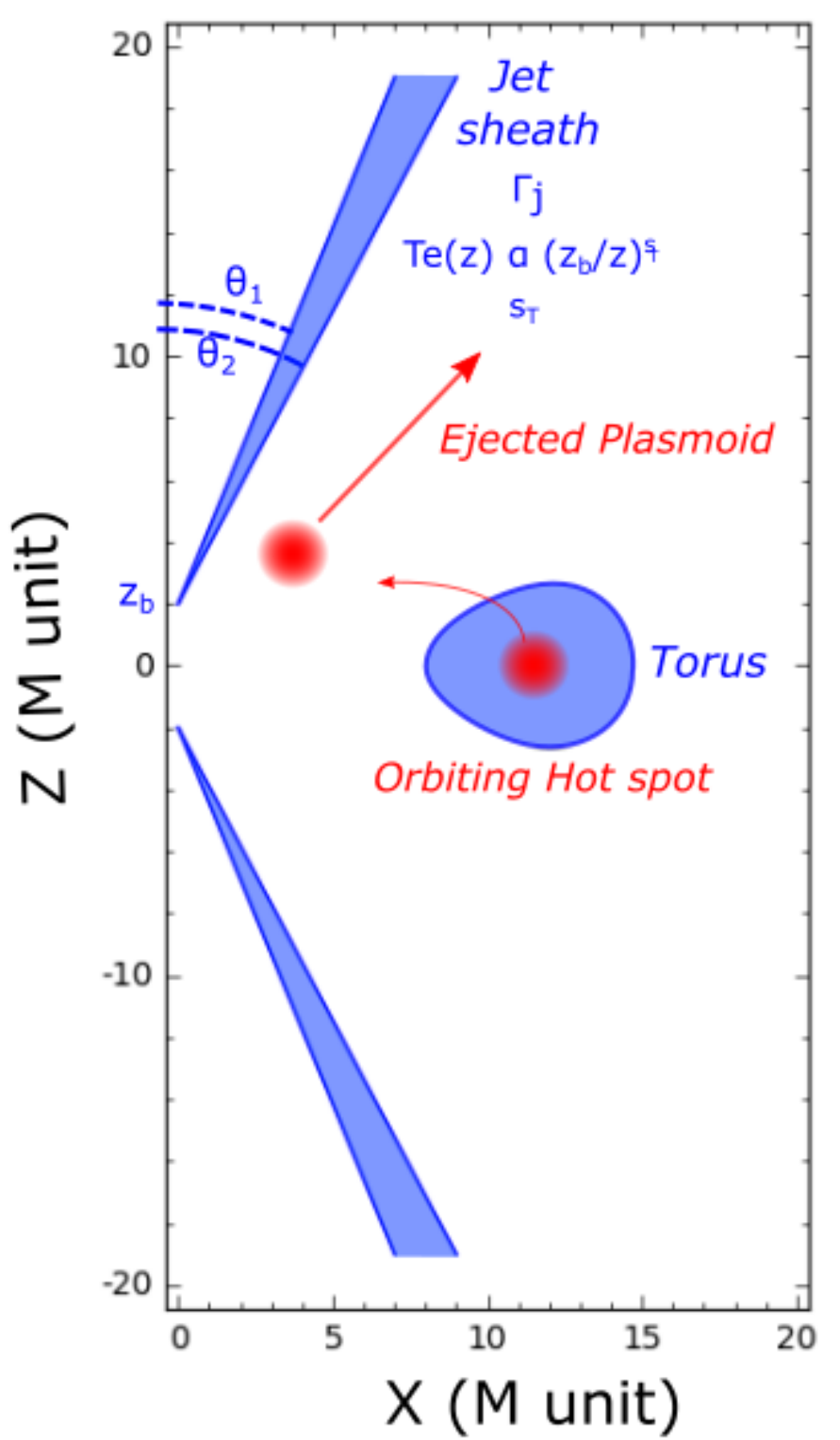}
    \caption{Scheme of the torus-jet model for the quiescent state in blue and flares in red. Two trajectories are considered for the flare, which can either rotate in the torus (hot spot model) or be ejected along the jet sheath (plasmoid model). The jet is parametrized by the angles $\theta_1$ and $\theta_2$ that describe the angular opening of the radiation-emitting sheath, by the base height $z_b$, the constant Lorentz factor $\Gamma_j$, and the temperature power-law index $s_T$. The jet is symmetrical with respect to the equatorial plane, and axisymmetric.}
    \label{fig:torus-jet model}
\end{figure*}

\section{Quiescent flow impact on astrometry: shifting and rotating the orbit}\label{sec:hot spot+ jet}

\citet{Gravity2018} used a hot spot model in an equatorial circular orbit to fit the astrometry of three bright flares.
They considered a constant radiation flux from the emitting region orbiting the black hole to fit the orbital motion.
The effect of out-of-plane motion and orbital shear have also been studied by \citet{Gravity2020} to model the flares.
However, the impact of the quiescent radiation surrounding the hot spot was
not taken into account. The aim of this section is to show that taking into
account the quiescent radiation can lead to shifting and rotating the
orbit on sky. We note a 1-$\sigma$ difference between the center of the orbit of the hot spot and the center of mass derived from the orbit of S2 in~\citet{Gravity2018} which makes this shift marginal.

In this section, we will use a simplified
hot spot model that is sufficient to highlight the main effects of the
quiescent radiation. This simple model will also allow us to introduce the
most important relativistic effects at play, that were already studied in many
previous works \citep{2006MNRAS.367..905B, 2009ApJ...692..902H}. These reminders will be helpful when we turn to a
more complex hot spot model in the Sect.~\ref{sec:Plasmoid}, which is the main aim
of this paper.

\subsection{Simple hot spot + quiescent model for the flaring Sgr~A*}\label{sec:quiescent model}

The quiescent radiation of Sgr~A* is modeled by means of the torus-jet
model as derived in~\citet{torus+jet}, to which we refer for all details.
Figure~\ref{fig:torus-jet model} resumes the main features of the model.
The torus emits thermal synchrotron radiation, while the flux emitted by the jet
follows a $\kappa$ distribution (i.e. a thermal core with a
power-law tail).
The multi-wavelength spectrum of the quiescent Sgr~A* is well fit with this
model. The $\kappa$ distribution emission from the jets dominates at most wavelength except at the sub-mm bump where the flux comes mostly from the thermal disk.
We summarize the best-fit parameters in Table~\ref{tab:quiescent table params},
and the resulting best-fit quiescent spectrum is given in
Fig.~\ref{fig:quiescent+spectrum}. More details on the fitting
procedure are given in Appendix~\ref{ap:quiescent}.
With these parameters, the flux of the torus-jet model at $2.2\ \mu m$ is $1.1 $ mJy. It is in perfect agreement with the median quiescent dereddened flux provided by \citet{2020A&A...638A...2G} of $1.1 \pm 0.3$ mJy. At this wavelength, the torus is optically thin and its emission is negligible compared to the jet.
In the remainder of this paper, where we focus only on the infrared
band, we will thus neglect the torus and consider a pure jet quiescent model,
unless otherwise noted.

The only relevant features of our quiescent model for the rest of this
paper are the location of its infrared centroid and its NIR flux. As depicted in the right
panel of Fig.~\ref{fig:quiescent+spectrum}, the centroid of our jet-dominated
model lies very close to the mass center. We have checked that considering
a disk-dominated model only very marginally changes the position of the
quiescent centroid at low inclination (see the blue and green dots in the
left panel of Fig.~\ref{fig:jet_influence}). Our conclusions are thus not
biased by our particular choice of a jet-dominated quiescent model.

\begin{table}[ht]
    \centering
    \begin{tabular}{ l c r }
        \hline
        \hline
        Parameter & Symbol & Value\\
        \hline
        \textbf{Black Hole} & &\\
        mass [$M_\odot$] & $M$ & $4.297 \times 10^6$ \\
        distance [kpc] & $d$ & $8.277$\\
        spin & $a$ & $0$\\
        inclination [$\degree$] & $i$ & $20$ \\
        \hline
        \textbf{Torus} & & \\
        angular momentum [$r_g/c$] & $l$ & $4$\\
        inner radius [$r_g$] & $r_{in}$ & $8$ \\
        polytropic index & $k$ & $5/3$\\
        central density [cm$^{-3}$] & $n_e^T$ & $1.2\times 10^{9}$\\
        central temperature [K] & $T_e^T$ & $7\times 10^9$\\
        magnetization parameter & $\sigma^T$ & $0.002$\\
        \hline
        \textbf{Jet} & & \\
        inner opening angle [$\degree$] & $\theta_1$ & $20$\\
        outer opening angle [$\degree$] & $\theta_2$ & $\theta_1+3.5$\\
        jet base height [$r_g$] & $z_b$ & $2$\\
        bulk Lorentz factor & $\Gamma_j$ & $1.15$\\
        base number density [cm$^{-3}$] & $n_e^J$ & $3.5 \times 10^6$\\
        base temperature [K] & $T_e^J$ & $3 \times 10^{10}$\\
        temperature slope & $s_T$ & $0.21$ \\
        $\kappa$ index & $\kappa^J$ & $5.5$ \\
        magnetization parameter & $\sigma^J$ & (fixed) $1$ \\
        \hline
        \hline
    \end{tabular}
    \caption{Best fit parameters of the torus+jet quiescent model. We keep the same geometrical parameters, bulk Lorentz factor and $\kappa$-index as \citet{torus+jet} and we fit the base number density, base temperature and temperature slope of the jet considering the correction (see bellow) and the new value of the jet magnetization parameter. The parameters of the torus are unchanged.}
    \label{tab:quiescent table params}
\end{table}

\begin{figure*}[ht]
    \centering
    \includegraphics[width=17cm]{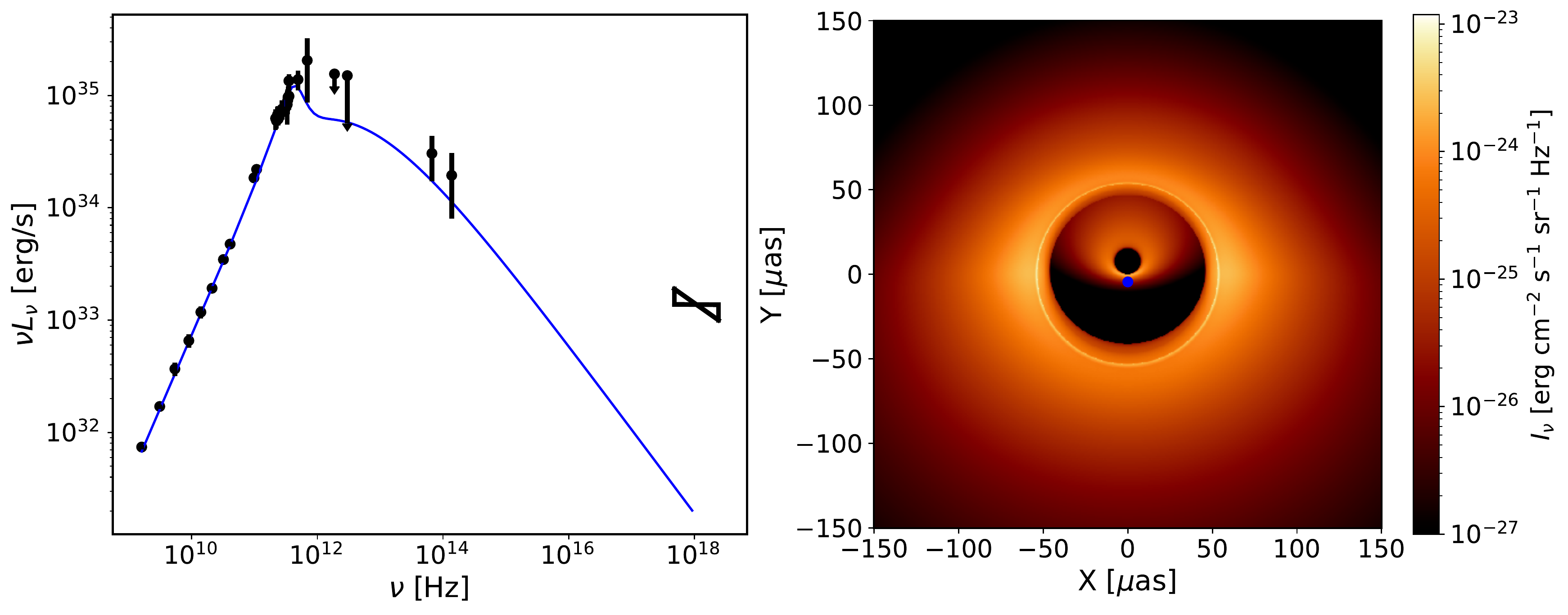}
    \caption{\textbf{Left:} Spectrum associated to the best-fit of the torus-jet model (see Table~\ref{tab:quiescent table params}) for the quiescent state of Sgr A* ($\chi^2_{red}=0.91$ with ndof=27). The data are taken from \citet{2015ApJ...802...69B} for $\nu < 50$ GHz, \citet{2015A&A...576A..41B} for the 2 points around 100 GHz, \citet{2016A&A...593A..44L} for the 492 GHz point, \citet{2006JPhCS..54..354M} for the 690 GHz point, \citet{2018ApJ...862..129V} for the far infrared upper limits, \citet{2018ApJ...863...15W} for the mid infrared data, and \citet{2001Natur.413...45B} for the X-ray bow-tie. We note that as in \citet{torus+jet}, the X-ray data are not fitted as we do not take into account bremsstrahlung nor Comptonized emission. \textbf{Right:} Best-fit image at $2.2$ $\mu m$ of the torus-jet model with a field of view of $150$ $\mu as$ seen with an inclination of $20 \degree$ and a Position Angle of the Line of Nodes (PALN) of $\pi$ rad. The color bar gives the values of specific intensity in cgs units in log-scale. The outer region emission comes from the backward jet's part while the emission close to the center comes from the forward part of the jet. The centroid of the jet is represented by the blue dot at $\sim$($0,-2.2$).}
    \label{fig:quiescent+spectrum}
\end{figure*}

The hot spot model is composed of a plasma sphere of radius 1 $r_g$ (fixed) with a uniform but time-dependent $\kappa$-distribution for the electrons. The emissivity $j_\nu$ and absorptivity $\alpha_\nu$ coefficients depend on the density, temperature, and magnetic field which we considered uniform. We use the fitting formula of \citet{2016ApJ...822...34P} to compute these coefficients. The typical light curve of a flare is characterized by a phase with increasing flux and one with decreasing flux. We model this behavior by a Gaussian time modulation on the density and temperature as follows

\begin{equation}
    n_e(t)=n_e^{hs}\ exp \left( -0.5 \times \left(\frac{t-t_{ref}}{t_{\sigma}}\right)^2\right),
    \label{eq:ne(t)}
\end{equation}
\begin{equation}
    T_e(t)=T_e^{hs}\ exp \left( -0.5 \times \left(\frac{t-t_{ref}}{t_{\sigma}}\right)^2\right)
\end{equation}

where $t_{\sigma}$ is the typical duration of the flare. As $n_e$ varies over time (Eq.~\ref{eq:ne(t)}), the magnetic field strength also varies since we set a constant
magnetization $\sigma = B^2 / 4 \pi m_p c^2 n_e$.

Contrary to \citet{Gravity2020}, we keep the circular equatorial orbit of \citet{Gravity2018} as we assume that the hot spot is formed in the equatorial plane and we do not take into account any shearing effect and assume a constant spherical geometry of the hot spot. We summarize all the input parameters of the hot spot in Table~\ref{tab:hotspot table params}.

\subsection{Shifting the orbit on sky}\label{sec:jet_influence}

\begin{figure*}[ht]
    \centering
    \includegraphics[width=17cm]{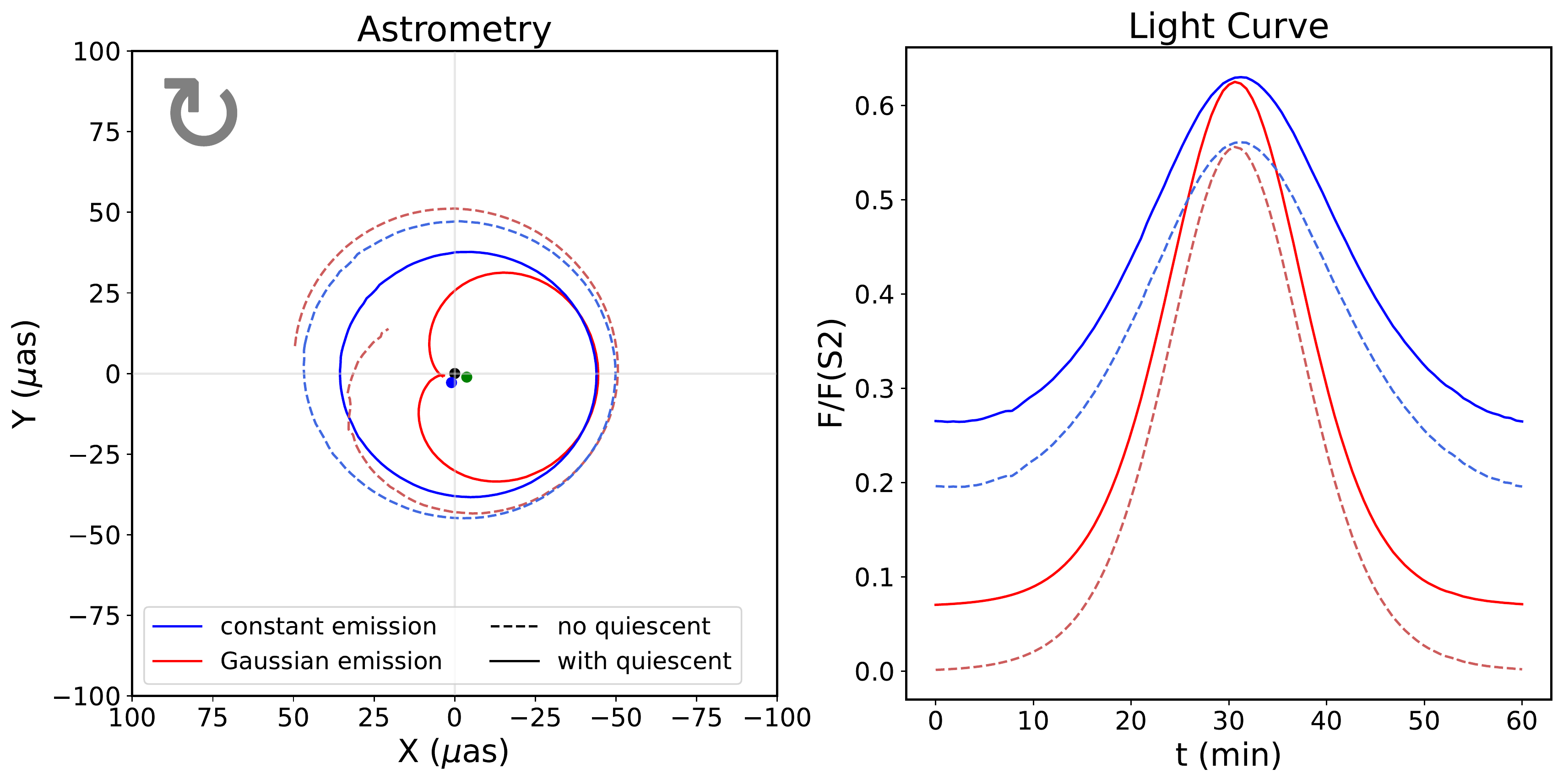}
    \caption{Astrometry (\textbf{left}) and light curves (\textbf{right}) of the hot spot - jet model with two values for the quiescent state corresponding to \textit{no quiescent} (dashed lines) and the \textit{with quiescent state} (full lines). In shade of blue, the hot spot has a nearly constant emission ($t_\sigma >> t_{orbit}$). The effect of beaming is reflected in the light curves. In shade of red, the hot spot has a Gaussian time emission with $t_\sigma=30$ min. The parameters of the hot spot are listed in Table~\ref{tab:hotspot table params}. We synchronise the maximum of beaming and the intrinsic maximum of the Gaussian modulation. The black, blue and green dots in the left panels represent the position of Sgr A*, the jet's centroid and the disk's centroid respectively.}
    \label{fig:jet_influence}
\end{figure*}

Figure~\ref{fig:jet_influence} shows the impact of taking into
account the quiescent radiation on the astrometry of the flare,
considering the trivial case of a constant-emission hot spot,
as well as the varying-emission hot spot introduced in
section~\ref{sec:quiescent model}.

Obviously, whether or not the hot spot intrinsic emission varies,
the first effect of adding a quiescent radiation is to shrink the
orbit's size, because the overall centroid is moved towards the quiescent
radiation's centroid, which always lies close to the mass center.

A slightly less obvious effect is that, when the hot spot
emission varies in time, the orbit can shift
in the plane of sky, and no longer be centered at the center-of-mass location.
This is clearly apparent on the solid-red orbit of the left panel of
Fig.~\ref{fig:jet_influence}. This is simply due to the time variation
of the intensity ratio between the quiescent and the hot spot
radiation. At early and late times, the hot spot has a weaker emission than
the quiescent component, and the overall centroid coincides with the quiescent centroid.
As the hot spot emission increases and dominates,
the overall centroid will be driven towards it.
Such a shift between the astrometric
data and the center-of-mass position is visible at 1-$\sigma$ significance in the
the~\citet{Gravity2018} data.

We note another non-trivial effect appearing in the
varying-emission hot-spot orbit without any quiescent radiation
(red-dotted orbit in Fig.~\ref{fig:jet_influence}).
The orbit is not closing, due to the time delay between the
primary and secondary images. Indeed, at the end of the simulation,
the flux from the secondary image is intrinsically higher than the primary
(the emission times of the primary and secondary are different),
and is amplified by the beaming effect.
When the centroid is computed, the secondary image has a larger impact at this time than before, resulting in a closer centroid position relative to the black hole.
This astrometric impact of the secondary image was already discussed by~\citet{2009ApJ...692..902H}.

\begin{table}[ht]
    \centering
    \begin{tabular}{l c r}
        \hline
        \hline
        Parameter & Symbol & Value\\
        \hline
        \textbf{Hot spot} & & \\
        number density max [cm$^{-3}$] & $n_e^{hs}$ & $1.05\times 10^7$\\
        temperature max [K] & $T_e^{hs}$ & $9.03 \times 10^{10}$\\
        time Gaussian sigma [min] & $t_\sigma$ & $30$\\
        magnetization parameter & $\sigma^{hs}$ & $0.01$\\
        $\kappa$-distribution index & $\kappa^{hs}$ & $5$\\
        orbital radius [$r_g$] & $R^{hs}$ & $9$\\
        initial azimuth angle [$\degree$] & $\varphi_0^{hs}$ & $90$\\
        Position Angle of the Line of Nodes [$\degree$] & $\Omega$ & $160$\\
        \hline
    \end{tabular}
    \caption{Summary of parameters of the hot spot model. We note that we used the maximum number density and temperature of the jet best-fit in Table~\ref{tab:quiescent table params} as reference and scale them for the hot spot by a factor $3.01$.}
    \label{tab:hotspot table params}
\end{table}

\begin{figure*}
    \centering
    \includegraphics[width=17cm]{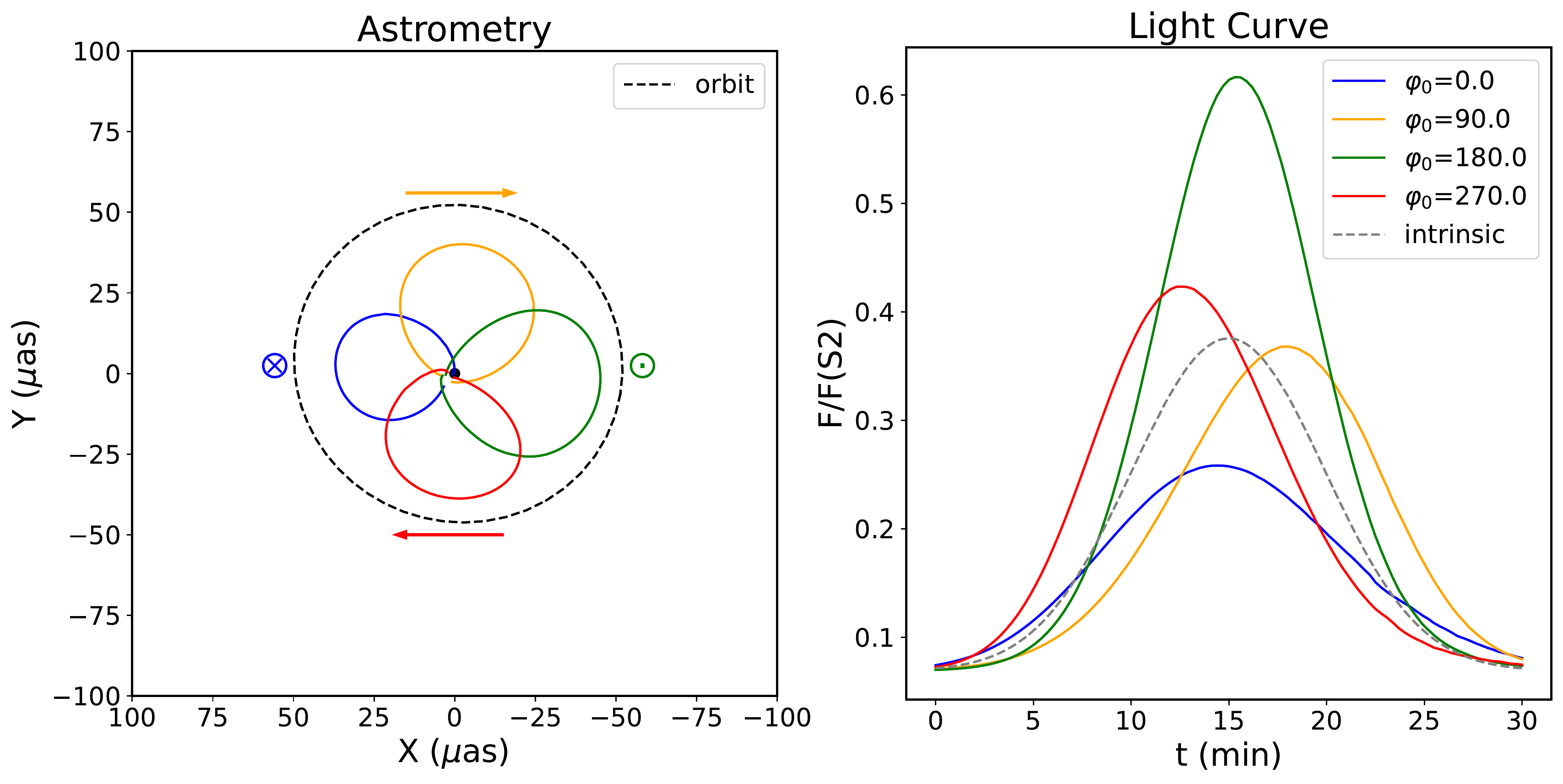}
    \caption{Astrometry (\textbf{left}) and light curves (\textbf{right}) of the hot spot - jet model for four initial azimuthal angle $\varphi_0$ of $0 \degree$ in blue, $90 \degree$ in orange, $180 \degree$ in green and $270 \degree$ in red. The dashed black line shows the primary image centroid track with no quiescent jet  (clock-wise). The jet dominates the beginning and the end of the flares. The observed centroids thus start and end close to the one of the jet. The apparent orbits rotate around the latter with $\varphi_0$ as the maximum of emission occurs at different $\varphi$. The Gaussian modulation which has a typical duration of $t_\sigma = 15$ min (grey dashed line; which is the same for the four the curves) is affected by relativistic effects. For negative X (right part of the astrometry), the beaming, combined with relativistic Doppler effect, amplify the flux from the hot spot while in the positive X (left part of the figure), they lower it. The black dot in the left panels represents the position of Sgr A*.}
    \label{fig:multi_phi0}
\end{figure*}

\subsection{Rotating the orbit on sky}\label{sec:time_variability}

It is not an easy task to disentangle the intrinsic time variability
of the hot spot from the variability due to the relativistic beaming effect.
Figure~\ref{fig:multi_phi0} illustrates the impact on astrometry and light curve
of playing with the relative influence between the intrinsic and beaming-related
variability. Here, we simply change the initial azimuthal coordinate
$\varphi_0$ of the hot spot along its orbit, in order to change the dephasing
between the time of the maximum intrinsic emission ($t=t_\mathrm{ref}$) which is fixed,
and the time of the maximum constructive beaming effect (when the hot spot
moves towards the observer). The orbit rotates around the quiescent centroid
following the variation of $\varphi_0$ (left panel of Fig.~\ref{fig:multi_phi0}). 
The light curve is also strongly affected, reaching much brighter levels when the intrinsic
emission maximum is in phase with the constructive beaming effect.

Here we show that the quiescent state of Sgr A* can have significant impact on the observed astrometry by shrinking the apparent orbit, creating a shift between the center of the latter and the position of the mass center. One should have these effects in mind for the comparison to the flare data at the end of the following section.

\section{Plasmoid model from magnetic reconnection}\label{sec:Plasmoid}

In this section, we develop a semi-analytical hot-spot-like
model in order to interpret the rise and decay of Sgr~A* flares, thus going one step further with respect to the model we used in section~\ref{sec:hot spot+ jet}, where a Gaussian modulation of the emission is enforced without physical motivation.

Black hole magnetospheres naturally lead to the development of equatorial current sheets corresponding to a strong spatial gradient of the magnetic field which changes sign at the equator~\citep{Komissarov04, Komissarov07, Parfrey19, Ripperda20}. Such a configuration results in magnetic reconnection, i.e. a change of the topology of the field lines forming X points~\citep{Komissarov05, Loureiro07, Sironi14}. This process is intrinsically non-ideal and thus can only be captured either by resistive MHD or kinetic simulations.
For suitable values of the magnetic diffusivity, the reconnecting current sheet can break into chains of plasmoids, i.e. magnetic islands separated by X points~\citep{Loureiro07, Parfrey19, Ripperda20, 2021MNRAS.502.2023P}. 

The reconnection rate (i.e. the typical rate at which magnetic energy is dissipated into particle kinetic energy) is equal to the ratio $v_\mathrm{rec}/v_\mathrm{out}$ with $v_\mathrm{rec}$ the velocity of matter injected
into the reconnection region, and $v_\mathrm{out}$ the bulk outflow velocity
of particles accelerated by the reconnection event.
The outflow velocity is of the order of
the Alfven speed, $v_\mathrm{out} \approx v_A$,
which is itself of the order of the speed of light,
$v_A \approx c$, for strongly magnetized environments.
The reconnection rate has been shown to be rather independent of the details of the chosen parameters.
For PIC simulations, it lies around $10\%$, i.e. $v_\mathrm{rec,PIC}\approx0.1v_A\approx0.1c$,
for magnetized collisionless plasmas~\citep{Sironi14, Werner18, 2015ApJ...806..167G}, which are the typical conditions in the inner flow surrounding Sgr A*\footnote{It is likely that the accretion flow surrounding Sgr A* is in a Magnetically Arrested Disk~\citep[MAD, see][]{Narayan03} regime, i.e. with strong poloidal magnetic fields in the inner regions~\citep{Gravity2018, Dexter20a}.}.
GRRMHD simulations point towards a slower 
rate of around $1\%$, so that $v_\mathrm{rec,MHD}\approx 0.01c$ (see the discussion in \citealt{Ripperda21}), but this applies to collisional environments, thus less similar to Sgr~A* vicinity. 

Fresh plasma flows into the current sheet at the reconnection rate $v_\mathrm{rec}$, is accelerated by the electric field generated in the current sheet, usually giving rise to power-law energy distributions of electrons~\citep{Sironi14, Werner18}. Inside the current sheet, the particles get trapped in the plasmoids which act as particle reservoirs~\citep{Sironi14} which can merge in a macroscopic magnetic island, that is, a large plasmoid. 
In \citet{Ripperda21}, magnetic flux dissipation through reconnection last for $\sim 100 r_g/c$ $\sim 30$ min and the resulting hot spot orbits for $\sim 500 r_g/c$ $\sim 150$ min before it disappears by losing its coherence through interaction with the surrounding flow.

In the global PIC simulation of \citet{Cerutti}, the authors study magnetic reconnection in the sheath of relativistic jet working with magnetic field loops coupling the BH to the accretion disk. The resulting plasmoids evolve off-plane, propagate away from the BH and are prone to merge with each other to form macroscopic plasmoids susceptible to radiate high amounts of energy under the form of non-thermal radiation. The underlying mechanism, first described by \cite{Uzdensky05} and \cite{2005A&A...441..845D}, relies on the accretion of poloidal magnetic field loops onto a spinning BH. Once the inner footpoint of the loop reaches the BH ergosphere, the magnetic field line experiences strong torques due to the frame dragging effect while its other footpoint on the disk rotates at the local Keplerian speed. Thus, the toroidal component of the magnetic field quickly grows in the innermost regions, propagates upstream along the field line and leads to the opening of the magnetic loop above a certain magnetic loop size. On the outermost closed magnetic field line (called the separatrix), a Y-point appears where plasmoids form and flow away along an inclined current sheet above the disk (Fig.~\ref{fig:reconnection_sketch}).
In the PIC simulations of \cite{Cerutti}, a cone-shaped reconnecting current sheet forms where vivid particle acceleration takes place. Electrons and positrons pile up into outflowing plasmoids where they cool through synchrotron radiation.
This topological configuration where some magnetic field lines anchored in the disk close within the event horizon is coherent with what is seen in resistive GRMHD simulations during the short episodes of flux repulsion which separates different accretion regimes. During approximately 100 rg/c, these simulations show an essentially force-free funnel surrounded by merging plasmoids formed in the jet sheath and in the equatorial plane \citet{Ripperda21,Chashkina+21}.

\subsection{Plasmoid model from magnetic reconnection}
\label{sec:plasmoid_model}

The aim of this section is to develop a semi-analytic large plasmoid model
(which will be named \textit{plasmoid model}),
inspired from the reconnection literature reviewed above. The interest of
such a model, compared to state-of-the-art GRMHD or GRPIC modeling,
is twofold:
\begin{itemize}
    \item it allows to remain as agnostic as possible regarding
    the physical conditions close to Sgr~A* and encapsulate into
    a single model a large parameter space;
    \item it allows to perform simulations within a limited
    computing time, allowing to explore the large parameter space and compare to astrometric and photometric data.
\end{itemize}
Our hope is that such a model can not only be fed with the
results of more elaborated simulations, but also bring constraints to these
simulations by determining what features of the modeling are important
in order to explain the data.

The main features of our model are
illustrated in Fig.~\ref{fig:reconnection_sketch}.
inspired by the recent GRPIC results of~\citet{Cerutti}.
Here, we consider a single plasmoid, which is modeled 
as a sphere of hot plasma with a constant radius. 
This macroscopic plasmoid is understood as the
end product of a sequence of microscopic plasmoid 
mergers. The spherical geometry is chosen only for
simplicity, given that current data are certainly unable to make a difference between various geometries. 

\subsubsection{Plasmoid motion}\label{ap:orbit}
We consider that the magnetic reconnection event occurs close to the black hole, and the resulting plasmoid is ejected along the jet sheath \citep{Ripperda20, Cerutti}. Thus, we define a conical motion (as in \citet{Ball_model}) defined by a constant polar angle $\theta=\theta_0$ and the initial conditions $r_0$, $\theta_0$, $\varphi_0$, $v_{r0}$, and $v_{\varphi0}$ . The subscript $0$ reflects the initial value of a given parameter in Boyer-Lindquist coordinate. As in \citet{Ball_model}, we set a constant radial velocity $v_r=v_{r0}$ and the azimuthal velocity is defined through the conservation of the Newtonian angular momentum

\begin{equation}
    v_\varphi(t) = v_{\varphi0}\frac{r_0^2}{r(t)^2}.
    \label{eq:phi_dot}
\end{equation}

The azimuthal angle is obtained by integrating the previous Eq.~\ref{eq:phi_dot}
\begin{equation}
    \varphi(t) = \varphi_0 + r_0^2 \frac{v_{\varphi0}}{v_r} (\frac{1}{r_0} - \frac{1}{r(t)}).
\end{equation}

In the GRPIC simulations performed by \citet{Cerutti}, the plasmoids are formed in the vicinity of the black hole at the Y-point and are ejected into the black hole magnetosphere, we thus restrict our study to $v_r > 0$.

An important feature of our model is the fact that the initial
azimuthal velocity of the plasmoid is naturally super-Keplerian.
Indeed, the Y-point, from which the plasmoid is generated, is anchored to the equatorial plane of the accretion
flow through the separatrix field line. So it will typically rotate at the
Keplerian speed corresponding to the foot point of the line, thus at
a velocity higher than the Keplerian velocity corresponding to the
initial cylindrical radius of the plasmoid.

\subsubsection{Growth and cooling phases}
We consider two phases in the lifetime of the plasmoid that aim at modeling the ascending and descending phases of the observed flare light curves:
\begin{itemize} 
	\item during the \textit{growth phase}, which lasts a total time $t_\mathrm{growth}$, the plasmoid continuously receives fresh accelerated particles at a constant rate resulting from the merging of microscopic plasmoids from reconnection into our large plasmoid which mix with "old" electrons cooled by synchrotron radiation. The growth time $t_\mathrm{growth}$ corresponds to the lifetime of the reconnection engine, i.e. the duration of magnetic flux dissipation; 
    \item after $t=t_\mathrm{growth}$, the plasmoid enters the \textit{cooling phase}: we assume that magnetic reconnection is quenched and plasmoids no longer merge so injection of fresh plasma stops and the plasmoid cools by emitting synchrotron radiation. We neglect particle escape and adiabatic losses.
\end{itemize}

The duration of the growth phase is set both by the reconnection rate and the speed at which magnetic flux is advected by the accretion flow into the current sheet. In \citet{Parfrey+15}, the accretion of successive magnetic loops of opposite polarity activates this process, with typical duration of transition of the order of 100$r_g/c$. This duration is representative of the dissipation of the magnetic flux of one magnetic loop which is set by both the size of the loop and the accretion speed, that the authors fix to 2$r_g$ and $c/200$ respectively, and the reconnection rate, fixed by the prescribed resistivity. Resistive GRMHD simulations of magnetically arrested disks \citep{Narayan03} bring support to these values \citep[e.g.][]{Ripperda20} but fail at reaching reconnection rates realistically high \citep{Bransgrove21}. In the more ab initio PIC simulations of \citet{Cerutti}, the reconnection rate is more realistic \cite[$\sim 10\%$,][]{Sironi14, Werner18} but due to the high computational cost of the simulations, they did not work over duration long enough to model the inward drift of the magnetic footpoints on the disk. As a consequence, the reconnection rate is accurate but the fueling magnetic flux is artificially steady and act as an infinite reservoir over the $\sim$200$r_g/c$ covered by the simulation. A coupling between GRMHD, force-free and PIC simulations to jointly describe the disk, the corona and the current sheet respectively is still missing. In this context, we considered duration $t_\mathrm{growth}$ of the growth phase of the order of 100$r_g/c$, corresponding to a typical episode of magnetic flux dissipation set by the two rates at which magnetic flux is advected into the current sheet and dissipated by magnetic reconnection.

\begin{figure}
     \resizebox{\hsize}{!}{\includegraphics{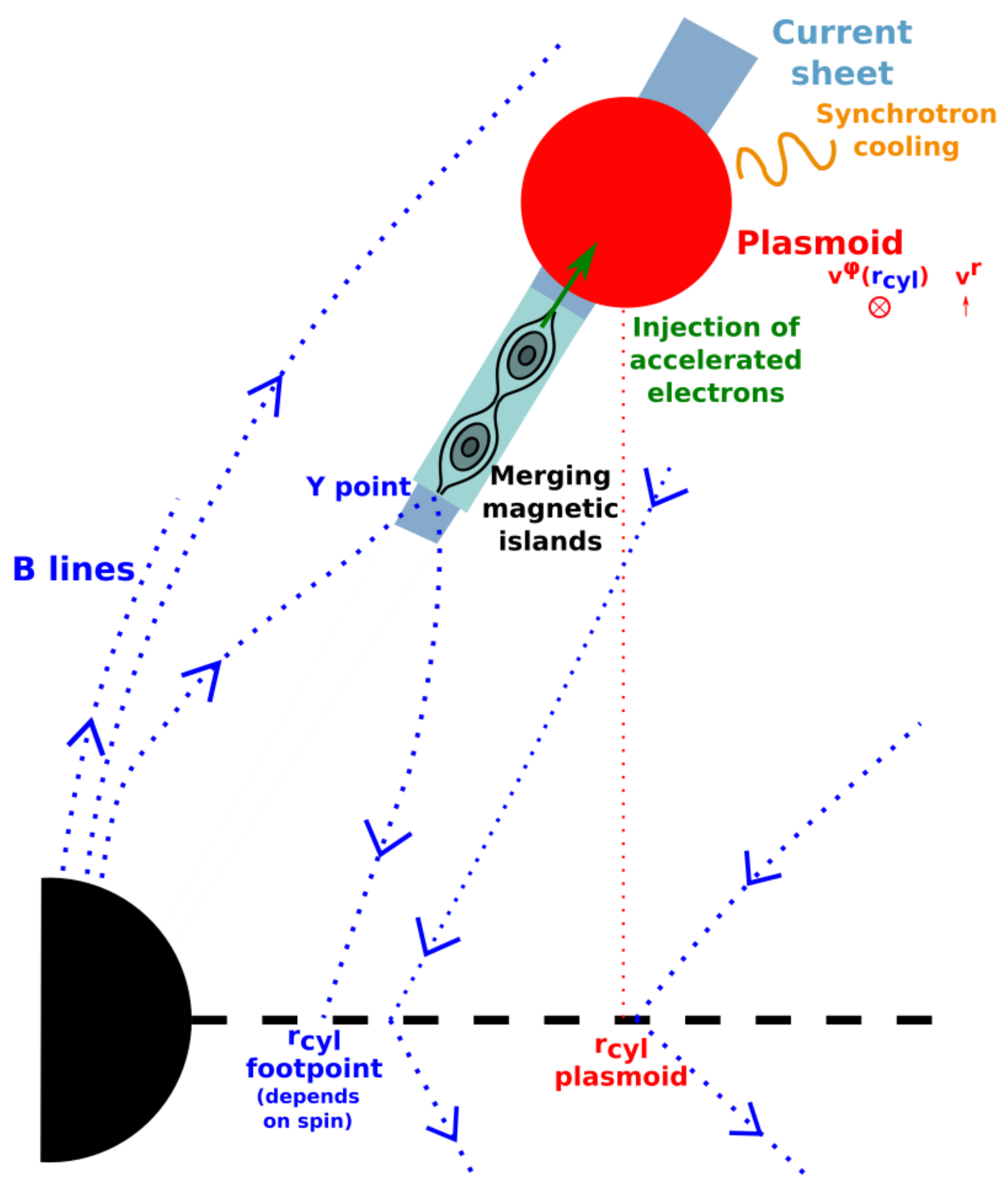}}
    \caption{Sketch of magnetic reconnection in the black hole magnetosphere as shown by \citet{Cerutti} on which our plasmoid model is built. There are three types of magnetic field lines: the ones threading the event horizon which goes to infinity, the ones anchored in the disk which also go to infinity, and the separatrix which link the disk and the black hole event horizon. The latter form a Y-point and a current sheet where chain of plasmoids are formed. Here we model a single plasmoid as the result of multiple mergers.}
    \label{fig:reconnection_sketch}
\end{figure}

\subsubsection{Evolution of the electron distribution}
Next, we prescribe the emission/absorption mechanism in the plasmoid. Most studies use chosen electron distributions, with analytical prescriptions for their evolution at best. \citet{Ball_model} use a fixed thermal distribution with a linear increase of the number density for the rising part of the light curve and a decrease of the temperature following Eq.~\ref{eq:elenevol} for the cooling. \citet{2022MNRAS.511.3536S} use a kappa distribution with an exponential cutoff and a synchrotron cooling break for the plasma emission generating X-ray flares. While their evolution of the plasma parameters (number density, temperature, magnetic field) is more elaborate than in our model,
their approximation is valid only while injection and cooling are balanced.
When the injection stops, the shape of the electron distribution changes rapidly (see Fig. \ref{fig:ElecDistribEvol}). Here, we choose a different approach by evolving the electron distribution in the plasmoid by solving the kinetic equation
\begin{equation} \label{eq:kineticeq}
\dfrac{\partial N_\text{e}(\gamma,t)}{\partial t} = \dfrac{\partial}{\partial \gamma}\left(-\dot{\gamma}_\mathrm{syn}\, N_\text{e}(\gamma,t) \right) + Q_\text{inj}(\gamma),
\end{equation}
where $\gamma$ is the Lorentz factor of the electrons, $Q_\text{inj}$ is the injection rate and $N_e = \mathrm{d} n_e / \mathrm{d} \gamma$ is the electron number density distribution, using the EMBLEM code~\citep{dmytriiev2021}. The term 
\begin{equation}
\label{eq:kinetic}
-\dot{\gamma}_\mathrm{syn} N_\text{e} = \frac{4 \sigma_T U_B}{3 m_e c} (\gamma^2 - 1) N_e
\end{equation}
of the right hand side describes the synchrotron cooling of the plasmoid particles, with $U_B = B^2/(8 \pi)$.
In our approach, we do not model the details of the magnetic reconnection process but instead describe the supply of freshly accelerated particles to the plasmoid by magnetic reconnection. Therefore, for the injection rate $Q_{\text{inj}}(\gamma)$ in Eq.~\ref{eq:kineticeq}, we use the following expression, assuming a constant injection rate:
\begin{equation} \label{eq:injterm}
    Q_{\text{inj}}(\gamma) = \left\{
    \begin{array}{ll}
        \dfrac{4 \pi N_e^\kappa(\gamma)}{t_\mathrm{growth}} & \mbox{in \:the\:growth \:phase, } \\
        0 & \mbox{in \:the\:cooling \:phase,}
    \end{array}
\right.
\end{equation}
where $N_e^\kappa(\gamma)$ is the distribution of the injected particles which follows the kappa distribution i.e. a thermal core with a power-law tail following:
\begin{equation} \label{eq:kappadistribut}
    N_e^\kappa(\gamma) = \dfrac{N}{4\pi} \gamma (\gamma^2 - 1)^{1/2} \, \left( 1 + \dfrac{\gamma - 1}{\kappa \Theta_e} \right)^{-(\kappa+1)}
\end{equation}
with a normalization factor
$N=1/2 n_e (\kappa - 2) (\kappa - 1) \kappa^{-2} \Theta_e^{-3}$, where $n_e$ and $\Theta_e$ are the density and dimensionless temperature of the injected plasma. The index $\kappa$ is defined as 
\begin{equation}
\kappa = p+1 = A_p + B_p \tanh{(C_p\ \beta_b)} + 1
\end{equation}
where
\begin{equation}
    A_p=1.8 + 0.7/\sqrt{\sigma_b}, B_p=3.7\ \sigma_b^{-0.19}, C_p=23.4\ \sigma_b^{0.26},
\end{equation}
following~\citet{2018ApJ...862...80B, Werner18}, where $p$ is the powerlaw index of the non-thermal part of the distribution, $\sigma_b~\gg~1$ is the plasma magnetization of the accelerating site and $\beta_b \ll 1$ is the ratio of proton thermal pressure to magnetic pressure of the accelerating site. If the magnetization at the accelerating site satisfies $\sigma_b \geq 100$ \citep{2021A&A...650A.163C}, then $\kappa$ is in the range of [$2.8, 4.4$] depending on $\beta_b$. This implies that the spectral index $\alpha$ ($\nu F_\nu \propto \nu^\alpha$) is between $-0.5$ and $0.5$ which is in perfect agreement with the measured indices for flares \citep[Fig. 32 in ][]{2010RvMP...82.3121G}. We note that realistic values for the magnetization in the funnel region of Sgr A* can be orders of magnitude higher than $100$~\citep[see][]{Ripperda21} which result in a smaller parameter space for $\kappa$, closer to the low boundary.

The bounds of the electron Lorentz factor are chosen to satisfy $\gamma_\mathrm{min}=1$ and $\gamma_\mathrm{max}=10^6$~\citep[Eq.3 of][]{Ripperda21}. When solving the kinetic Eq.~\ref{eq:kineticeq}, we assume that the density of the plasmoid particles follows
\begin{equation}
    n_e(t) = \left\{
    \begin{array}{ll}
        n_e^\mathrm{max} \times t / t_\mathrm{growth} & \mbox{in \:the\:growth \:phase, } \\
        n_e^\mathrm{max} & \mbox{in \:the\:cooling \:phase.}
    \end{array}
\right.
\end{equation}
Such high maximum Lorentz factor is needed to also power X-ray flares with only synchrotron. However, \citet[][]{Ripperda21} suggest a lower maximum Lorentz factor $\gamma_\mathrm{max} \sim 10^4$ in the plasmoid as electrons cool during their travel time between the acceleration site and the later. Such lower value results in a marginally lower flux in NIR, as most of the emission at this wavelength comes from lower energy electrons, which can be compensated with a slightly higher maximum number density.

The temperature of the injected particles remains fixed in the growth phase, and we define a uniform and time-independent tangled magnetic field in the plasmoid. This is also a simplifying
assumption, and we intend to consider in future work
the impact of the magnetic field geometry on the
polarized observables.

The EMBLEM code does not only solve for the evolution of the electron distribution, it also provides the associated synchrotron emission and absorption coefficients of the plasmoid particles. We can thus compute an image of our plasmoid scenario by backwards integrating null geodesics 
in the Schwarzschild spacetime from a distant observer screen, and integrate the radiative transfer equation through the plasmoid by reading the tabulated emission and absorption provided by EMBLEM. This step is performed by means of the
\texttt{GYOTO}\footnote{\href{https://gyoto.obspm.fr}{https://gyoto.obspm.fr}} ray-tracing code \citep[see Appendix~\ref{sec:GYOTO} for details;][]{GYOTO, paumard_thibaut_2019_2547541}.
The input parameters that we used for the code are listed in Table~\ref{tab:inputparamemblem}. With these values of density and magnetic field strength, we obtain a magnetization inside the plasmoid of $\sigma_p \sim 10^{-2}$ from the end of the growth phase since neither the density nor the magnetic field evolve.

\subsection{Importance of evolving the electron distribution}

One of the most important aspects of our model is the
self-consistent evolution of the electron distribution function,
and corresponding radiative transfer, in the plasmoid. Here, we illustrate the importance of taking
into account the evolution of the electron distribution by comparing
our model with another reconnection plasmoid model inspired by \citet{Ball_model}.

\begin{figure*}
    \centering
    \includegraphics[width=17cm]{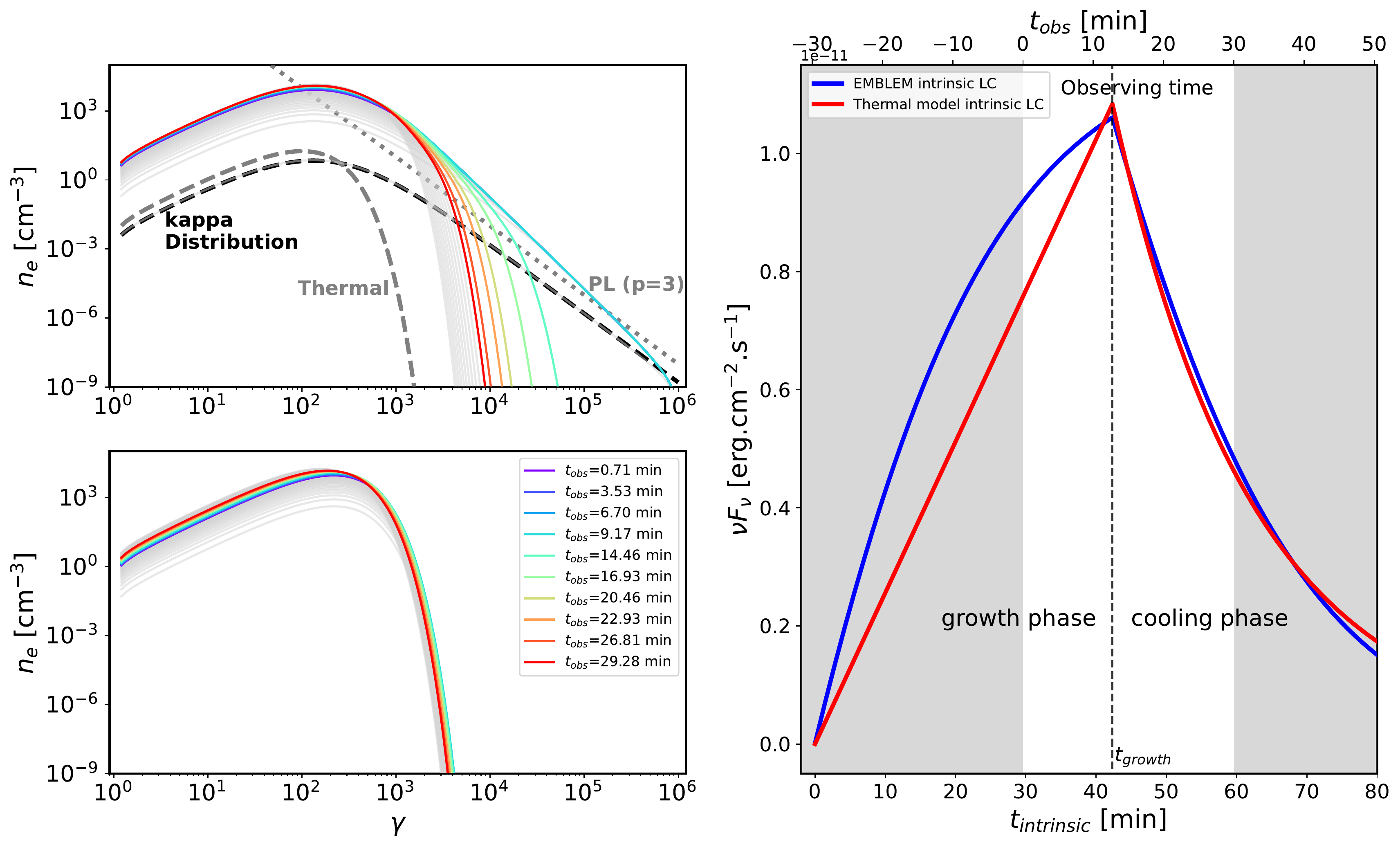}
    \caption{(\textbf{Top-left}) Evolution of the electron distribution in our model with \texttt{EMBLEM} at each observing time of the July 22, 2018 flare. The black dotted line correspond to the injected $\kappa$ electron distribution composed of a thermal core with a power law tail. The parameters used are listed in Table~\ref{tab:inputparamemblem}. (\textbf{Bottom-left}) Evolution of the electron distribution in the Thermal model inspired by \citet{Ball_model} at each observing time of the July 22, 2018 flare. The parameters used for this distribution are the same as in our model (listed in Table~\ref{tab:inputparamemblem}) but with the dimensionless temperature of $\Theta_e=109$ and the magnetic field of $B = 20$ G. (\textbf{Right}) Full intrinsic light curves of the two models. Note that in this panel we plot the light curve from the beginning of the growth phase while in the left panels we plot the distribution at the observed time of Fig.~\ref{fig:plasmoid_flare} ($t_{obs}=t_{intrinsic}-29.6$ min).}
    \label{fig:ElecDistribEvol}
\end{figure*}

\begin{figure*}
    \centering
    \includegraphics[width=17cm]{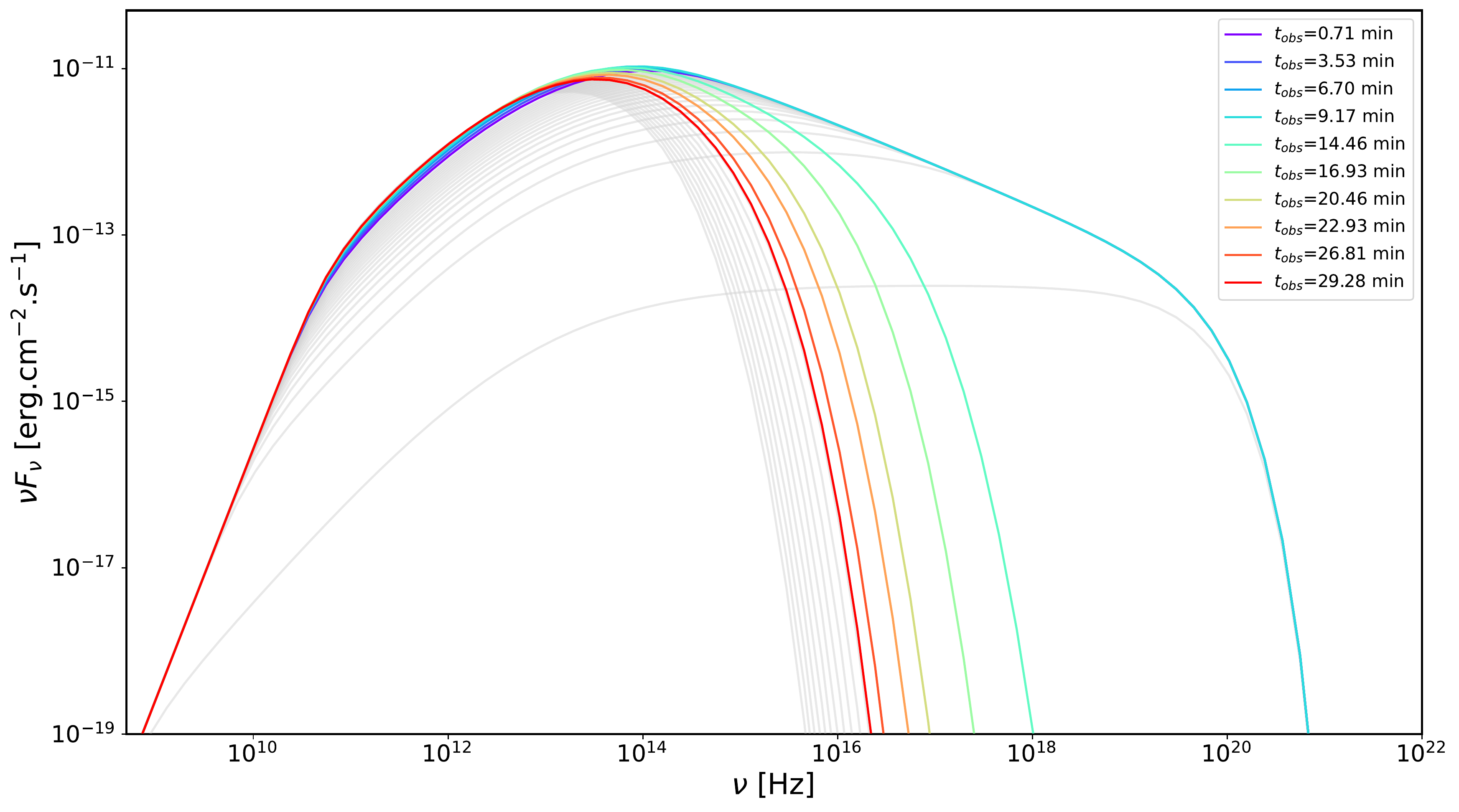}
    \caption{Intrinsic Spectral Energy Density (SED) evolution from radio to X-rays of our plasmoid model with the parameters listed in Table~\ref{tab:inputparamemblem}. Color code the time as in Fig.~\ref{fig:ElecDistribEvol}. Grey lines are the SED out of the observing time. The SED peak occurs in NIR with this set of parameters, but with a lower $\kappa$-index, i.e. a harder power law tail, the peak could reach X-rays. Here, X-rays flux drops quickly compare to the typical timescale of X-ray flares as we consider only synchrotron cooling and not Synchrotron-Self Compton (SSC). We note that the flux emitted by our plasmoid at 230 GHz is $\sim 0.1$ Jy which is the good order of magnitude of the variability associated to flares in sub-mm \citep{2022ApJ...930L..19W}.}
    \label{fig:sed-evolved}
\end{figure*}

We show in the top-left panel of Fig.~\ref{fig:ElecDistribEvol} the evolution of the electron distribution in our plasmoid model, for the parameters listed in Tables~\ref{tab:inputparamemblem} and \ref{tab:plasmoid_orbital_params}
(see Sect.~\ref{sec:ray-tracing} for details) and the associated spectral energy density (SED) in Fig.~\ref{fig:sed-evolved}. During the growth phase ($t_{obs}< 10$ min), for $\gamma > 10^3$ the distribution is stationary as the injection is balanced by the cooling. After the end of the growth phase, the shape of the distribution changes rapidly as there is only cooling left. We show in the right panel of Fig.~\ref{fig:ElecDistribEvol} the light curve obtained with our model (in red) and with a model inspired by \citet{Ball_model} who do not take into account the non-thermal electrons, i.e. using a thermal distribution, with a linear increase of the number density with a fixed temperature during the growth phase and an analytical prescription for the temperature decrease during the cooling phase using Eq.~\ref{eq:elenevol} and assuming $\Theta_e = \gamma /3$. While this model gives a similar intrinsic light curve as our model, the dimensionless temperature required is twice as high as ours ($\Theta_e=109$) with a magnetic field of $B=20$ G to cool faster lower energy electrons. The evolution of the distribution with this model is shown in the bottom left panel of Fig.~\ref{fig:ElecDistribEvol}. We do not need such high a temperature as most of the emission comes from high-energy electrons which are non-thermal in our model as suggested by PIC simulations \citep{2017ApJ...850...29R, Werner18, 2018ApJ...862...80B, 2021ApJ...922..261Z}. Our temperature could be even lower with a harder (i.e. lower) $\kappa$-index. The cooling of the electron distribution through synchrotron radiation is difficult to model properly and needs a kinetic approach as we do in our plasmoid model.

\subsection{Comparing GRAVITY 2018 flare data with our plasmoid model}\label{sec:ray-tracing}

\begin{figure*}
    \centering
    \includegraphics[width=17cm]{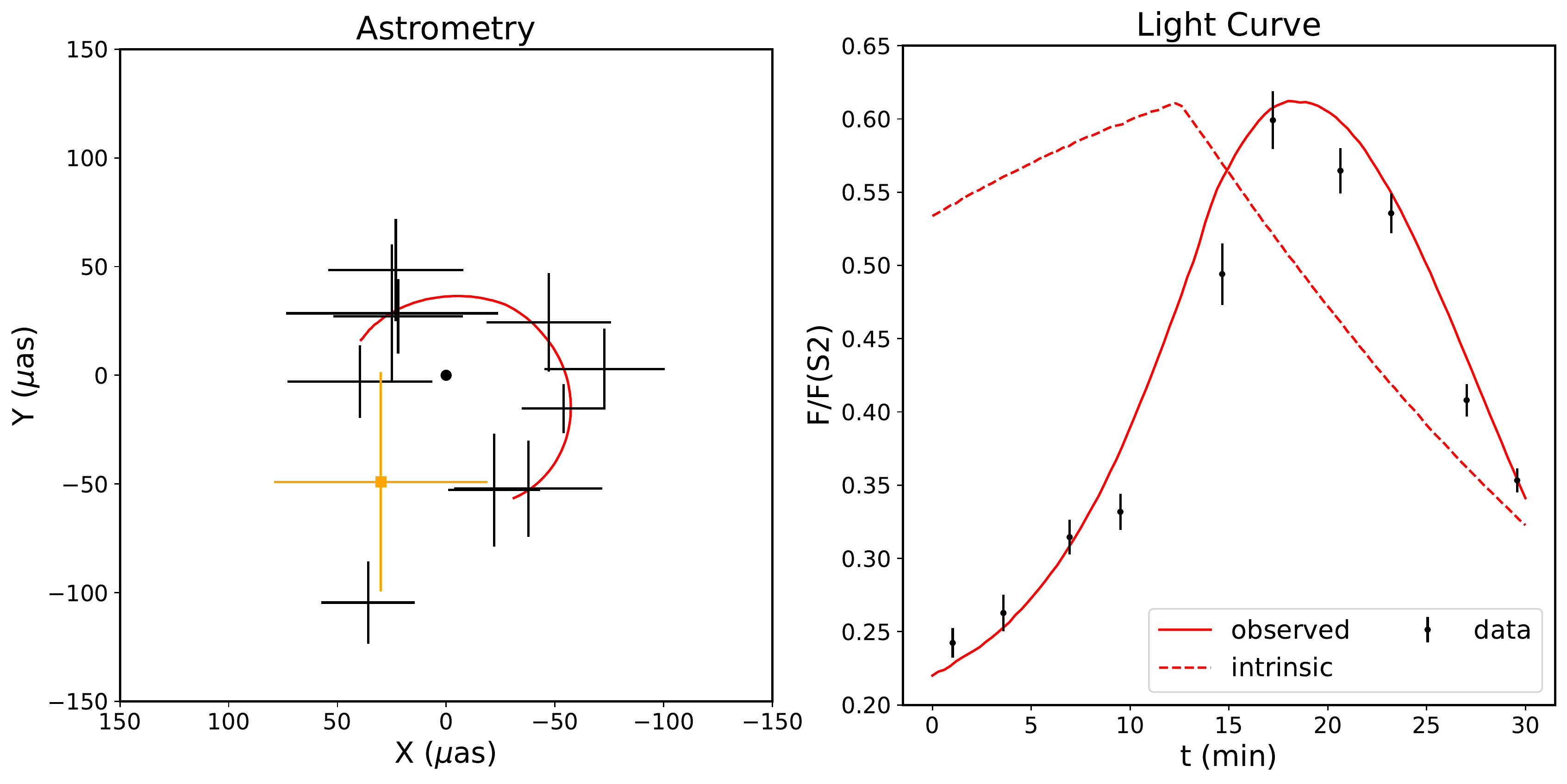}
    \caption{Data and plasmoid models of the flares from July 22, 2018. The left panels shows the astrometry of the flare while the right panel shows the observed (full line) and intrinsic (dashed line) light curves. The parameters of the model are listed in Tables~\ref{tab:inputparamemblem} and \ref{tab:plasmoid_orbital_params}. Note that this is not the result of a fit. The black dot in the left panels represents the position of Sgr A* in \texttt{GYOTO} and the orange cross represent the position of Sgr A* measured through the orbit of S2.}
    \label{fig:plasmoid_flare}
\end{figure*}

This paper aims at checking whether we can reproduce with our plasmoid model the general features of the observed light curve and astrometry of the July 22, 2018 flare reported by \citet{Gravity2018}. 
We show in Fig.~\ref{fig:plasmoid_flare} a comparison between the July 22, 2018 flare data observed by GRAVITY (in black) and our plasmoid model (red line) with the parameters listed in Tables~\ref{tab:inputparamemblem} and \ref{tab:plasmoid_orbital_params}.
For comparison, we show the intrinsic light curve (dashed line) obtained by removing all the relativistic effects (Doppler effect, beaming, secondary image). 

This comparison is not the result of a fit and was obtained by estimating the relevant parameters using simple physical arguments:
\begin{itemize}
\item The rise time and slope of the light curve
are mainly monitored by (i) the growth time, (ii) our
choice of linear evolution of the electron
density (which enters the injection function), (iii) the relativistic 
beaming effect, and thus (iv) the
initial azimuthal position of the plasmoid,
$\varphi_0$, which has a strong impact on 
beaming as illustrated in the right
panel of Fig.~\ref{fig:multi_phi0}.
\item The decaying part of the
light curve is
monitored by the synchrotron cooling time, thus by the magnetic
field strength, and by the beaming effect. 
\item The maximum of the light curve can be estimated
by means of an analytical formula that we derive in
Appendix~\ref{sec:plasmoid_analytic}.
This maximum depends mainly on the maximum number density $n_{e,max}$, as well as on the temperature and $\kappa$-index of the distribution. These parameters are degenerate and thus not constrained with only the NIR flare data. Nevertheless, GRMHD \citep{Dexter20a, Ripperda21, 2022MNRAS.511.3536S} and GRPIC simulations \citep{Cerutti} of magnetic reconnection suggest that the density in the plasmoid is higher than its close environment in the current sheet, of the order of the density at the base of the jet, close to the event horizon, but lower than in the disk. Still, the two remaining parameters ($\Theta_e$, $\kappa$) which describe the shape of the distribution are fully degenerate.
\item The initial position and velocity of
the plasmoid have a strong impact on
the astrometric trace on sky.
We guess the initial azimuthal velocity based on the following reasoning.
The Keplerian velocity of the plasmoid at its initial cylindrical radius is $v_{Kep} \sim 0.31c$ (for our choice of initial cylindrical radius given in Table~\ref{tab:plasmoid_orbital_params}, $r_{cyl} = 10.6~r_g$). However, as discussed in Sect.~\ref{sec:plasmoid_model}, our model naturally leads to a super-Keplerian
initial velocity to the plasmoid. Indeed, the plasmoid initial azimuthal
velocity is that of the footpoint of the separatrix
(see Fig.~\ref{fig:reconnection_sketch}).
Based on Fig. 8 of \citet{Cerutti}, we can determine the radius of the
footpoint, $r_{fp}$, of a separatrix giving rise to a Y point located at
a cylindrical radius of $\approx 10\,r_g$.
We find $r_{fp} = (4.7 \pm 0.5)\, r_g$, which translates in an orbital velocity $v_{\varphi,0}$ between $0.41c$ and $0.45c$. The upper bound of this interval compares
well with the July 22 flare data. 
We note that this estimate of the initial
azimuthal velocity is anchored in the
model of~\citet{Cerutti} and thus depends
on their choice of initial condition,
in particular on the initial profile of
their magnetic field.
\end{itemize}

We note that the fiducial values proposed in Tables \ref{tab:inputparamemblem} and \ref{tab:plasmoid_orbital_params} represent a set of parameters with values that are inspired by numerical simulations of reconnection which reproduce the key observational features of the July 22 flare data. This setup is not unique and is not the result of a fit. We let the exploration of the full parameters space (freeing some fixed/constrained parameters like maximum number density, growth time, inclination) to a future work. Nevertheless, our model disfavor low growth time ($t_\mathrm{growth} < 50 r_g/c$) for this particular flare.

Overall, our plasmoid model jointly describes the astrometry and the flux variation of the 22 July 2018 flare measured by~\citep{Gravity2018} for the first time, considering a model with a specific emission prescription. Magnetic reconnection is thus a viable scenario to explain Sgr A* flares.

\begin{table}
    \centering
    \begin{tabular}{lccc}
        \hline
        \hline
        Parameter & Symbol & Value\\
        \hline
        \textbf{Plasmoid} & & \\
        {magnetic field} [G] & $B_\text{p}$ & 15 \\
        {plasmoid radius} [$r_g$] & $R_{\text{p}}$  & $1$ \\
        {minimal Lorentz factor} & $\gamma_{\text{min}}$ & 1 \\
        {maximum Lorentz factor} & $\gamma_{\text{max}}$ & $10^6$ \\
        {kappa distribution index} & $\kappa$ & 4.0 \\
        {kappa distribution temperature} & $\Theta_e$ & 50 \\
        {maximum electron number density} [cm$^{-3}$] & $n_{\text{e,max}}$ & $5 \times 10^6$ \\
        {growth timescale} [$r_g/c$] & $t_\mathrm{growth}$ & 120 \\
        \hline
    \end{tabular}
    \caption{Input parameters of the \texttt{EMBLEM} code for the simulation of the electron distribution evolution. These parameters are used for the July 22 flare of \citet{Gravity2018}.}
    \label{tab:inputparamemblem}
\end{table}

\begin{table}
    \centering
    \begin{tabular}{lcc}
        \hline
        \hline
        Parameter & Symbol & July 22 \\
        \hline
        \textbf{Plasmoid} & &\\
        time in \texttt{EMBLEM} at zero observing time [min] & $t_{obs,0}^{emblem}$ & $29.6$\\
        initial cylindrical radius [$r_g$] & $r_{cyl,0}$ & $10.6$\\
        polar angle [$\degree$] & $\theta$ & $135$\\
        initial azimuthal angle [$\degree$] & $\varphi_0$ & $280$\\
        initial radial velocity [$c$] & $v_{r,0}$ & $0.01$\\
        initial azimuthal velocity [$c$] & $v_{\varphi,0}$ & $0.45$\\
        X position of Sgr A* [$\mu as$] & $x_0$ & $0$\\
        Y position of Sgr A* [$\mu as$] & $y_0$ & $0$\\
        PALN [$\degree$] & $\Omega$ & $160$\\
        \hline
    \end{tabular}
    \caption{Orbital parameters of the plasmoid model following a conical motion used for the comparison of the July 22, 2018 flares observed by \citet{Gravity2018}.}
    \label{tab:plasmoid_orbital_params}
\end{table}

\section{Limitations of our plasmoid model}\label{sec:limitation}

Our plasmoid model is vastly simplified with respect to the
complexity of realistic magnetic reconnection events in the
environment of black holes. We review here its main limitations:
\begin{itemize}
    \item[i.] We consider a single plasmoid while the instability
    of thin current sheets gives rise to a dynamic flow of
    merging magnetic islands. Our argument for this simplification
    is that the merging process is certainly very dependent on the 
    unknown initial conditions, and that the final, bigger
    and brighter product of the cascade is likely to dominate
    the observed signal;\\
    \item[ii.] The initial condition on the plasmoid's velocity
    is simply imposed for the radial motion, and based
    on a particular GRPIC model as regards the azimuthal motion;\\
    \item[iii.] The evolution of the plasma parameters (density, temperature, magnetic field) are chosen to be
    either constant or linear, so very
    simplified compared to a realistic 
    scenario. However, we consider that these evolutions are
    very likely to be strongly dependent on the initial conditions
    of the flow, so that they are weakly
    constrained;\\
    \item[iv.] The values of almost all the parameters except mass and distance of Sgr A* are poorly constrained. We choose a set of values which are reasonable according to simulations. Future work is needed to investigate the details of the parameter space.\\
    \item[v.] We model the plasmoid by a homogeneous sphere for simplicity from the circle plasmoid seen in 2D GRMHD \citep{Nathanail20,Ripperda20, 2021MNRAS.502.2023P} and PIC simulations \citep{2017ApJ...850...29R, 2018ApJ...862...80B, Werner18}. The 3D aspect of such plasmoid is cylindrical (flux ropes) both in GRMHD \citep{2021PhRvL.127e5101B, 2022MNRAS.513.4267N, Ripperda21} and PIC \citep{2021ApJ...921...87N, 2021ApJ...922..261Z} simulations.
    Thus, a realistic geometry of the flare source is likely more complex than in our model. We note that the exact geometry of the flare is not relevant as we only track the centroid position, as much the flare source is not too extended, and we consider tangled magnetic field. However, the coherence time
    of the structure might be shorter in 3D and might have an impact
    on the rise time of the light curve. Further 3D simulations studies are needed to better model the shape of the flux ropes and their evolution;\\
    \item[vi.] We neglect any shearing of the plasmoid and
    consider that it remains identical to itself throughout
    the simulation. Differential rotation is however likely to
    stretch the plasmoid over its orbit and destroy its coherence \citep{2009ApJ...692..902H, Gravity2020};\\
    \item[vii.] We consider a tangled magnetic field in the
    plasmoid and thus do not consider the impact of
    the magnetic field geometry on the observable. 
    The magnetic field geometry of the quiescent
    flow is likely to be ordered and vertical if
    Sgr~A* is strongly magnetized.
    The magnetic field in the plasmoid, which is our interest here, could be either helical (plasmoids) or vertical for large flux tubes \citep{Ripperda21}.\\
    \item[viii.] During a flare, the quiescent state can change in a non axisymmetrical way~\citep{Ripperda21}. This will push the centroid position of the quiescent further away from the center-of-mass location which will affect the offset. We choose to use a static and axisymmetric quiescent model during flare to avoid adding more degrees of freedom which would lead to higher degeneracies, rather than clearer constraints.\\
    \item[ix.] We choose a high maximum Lorentz factor $\gamma_\mathrm{max}=10^6$ to be able to power X-ray flares (but without any constrain for this study). However, high energy photons lead to pair-production and thus increase the number density in the plasmoid which we do not take into account.
\end{itemize}

Despite these many limitations, we consider that our model
is very interesting for fitting flare data, because it allows
to cover a much broader set of physical scenarios than 
more elaborate simulations that strongly depend on their
assumptions regarding the relevant physics and the initial
conditions.

\section{Conclusion and perspectives}\label{sec:discussion}

This paper is mainly focused at developing a new plasmoid model
for Sgr~A* flares, inspired by magnetic reconnection in black
hole environments. Our semi-analytic model allows to study a broad parameter space within
a reasonable computing time, thus being well suited for
data analysis.

Our model considers non-thermal electrons accelerated by magnetic reconnection and injected into a spherical large plasmoid. We evolve the electron distribution through a kinetic equation taking into account synchrotron cooling and particle injection at a constant rate. We show in Appendix~\ref{sec:Plasmoid_test} (Fig.~\ref{fig:EMBLEM_cooling}) the importance of taking into account the cooling of the electrons already in plasmoid during the growth phase. 
Our model also naturally accounts for a super-Keplerian velocity
of the flare source, through the dynamical coupling between the
plasmoid and the inner regions of the accretion flow through
magnetic field lines.
One of the main results of this paper is that for the first time we model the
astrometry and lightcurve of the flares measured by~\citep{Gravity2018} by explicit modeling of the emission zone.
 
Our conclusions regarding the three main points raised
in the introduction are the following:
\begin{itemize}
\item the marginally detected shift between the astrometric track
of~\citet{Gravity2018} and the center of mass might be due to the impact
of the quiescent radiation of the background accretion flow;
\item a dynamical coupling between the plasmoid and the inner
accretion flow through closed magnetic field lines might naturally account for the super-Keplerian
speed obtained by~\citet{Gravity2018};
\item in general, a large plasmoid due to magnetic reconnection in
a thin current sheet in the black hole magnetosphere is a reasonable
model to account for the main features of the~\citet{Gravity2018} observables.
\end{itemize}

Sect.~\ref{sec:ray-tracing} shows that the
temperature, density, and $\kappa$ parameters
of the plasmoid are degenerate.
This degeneracy might be removed by 
simultaneous observations of NIR and X-ray flares. 
Moreover, synchrotron cooling
leads to a translation of
the electrons from the NIR-emitting band
to the millimeter-emitting band,
which could explain the sub-mm flare and its time lag with respect to NIR. We thus intend to consider the
multi-wavelength properties of our plasmoid
model in future work, in order to better assess its ability to account for the complete flare data set of Sgr~A*.

A crucial recent observable of Sgr~A* flares
are the polarization QU loops~\citep[][]{Gravity2018,2020A&A...643A..56G,ALMA_polar}.
We also intend to study the polarized properties of our plasmoid model
and compare it to these recent constraints.

\begin{acknowledgements}
NA and FHV are very grateful to B. Cerutti, B. Crinquand, S. von Fellenberg, S. Gillessen, S. Masson, B. Ripperda, N. Scepi, and M. Wielgus for fruitful discussions.
This work was granted access to the HPC resources of MesoPSL financed by the Region Ile de France and the project Equip@Meso (reference ANR-10-EQPX-29-01) of the programme Investissements d’Avenir supervised by the Agence Nationale pour la Recherche.
\end{acknowledgements}

\bibliographystyle{aa}
\bibliography{biblio.bib}

\begin{appendix}
\section{Torus - Jet model for the quiescent state}\label{ap:quiescent}
We used the Torus-jet model of \citet{torus+jet}. Their jet model is restricted to an emitting sheath with an empty funnel in agreement with GRMHD simulations \citep[e.g.][]{2013A&A...559L...3M, 2017MNRAS.467.3604R, 2018A&A...612A..34D, 2019ApJS..243...26P}. In their model, \citet{torus+jet} define an opening and closing angle $\theta_1$ and $\theta_2$ respectively and a base height $z_b$ to define the geometry of the jet. The number density and the temperature are defined by their values at the base height of the jet ($n_e^J$ and $T_e^J$ respectively) and their profiles along the jet. The profile of the number density is fixed ($\propto r_{cyl}^{-2}$ with $r_{cyl}$ the projected radius in the equatorial plane) and the one of the temperature is set by the temperatures slope $s_T$ ($\propto z^{-s_T}$ with $z$ the height along the vertical/spin axis). The jet emits synchrotron radiation from a $\kappa$ electron distribution.

The torus is defined by its central density and temperature ($n_e^T$ and $T_e^T$ respectively). The profiles of these two quantities in the torus are governed by the polytropic index $k$ and its geometry. The latter is defined by the inner radius $r_{in}$ and the angular momentum $l$ but also on the metric (see \citet{torus+jet} for more details). Contrary to the jet, we consider that the electron distribution of the torus is purely thermal.

We use the same algorithm as in \citet{torus+jet} after the correction of a small technical issue leading to an overestimation of the number density and temperature. However, we change the choice of the magnetization parameter in the jet sheath. As illustrated e.g. by \citet{2019ApJS..243...26P}, the jet sheath, which corresponds to the dominating emission region of the jet, coincides with the transition between the highly-magnetized ($\sigma \gg 1$) funnel and the less-magnetized ($\sigma \ll 1$) main disk body. Consequently, we fix the magnetization to  $\sigma = 1$ in the emitting jet sheath, while \citet{torus+jet} used a low magnetization both in the jet and in the torus. Our choice leads to a smaller density in the jet sheath compared to \citet{torus+jet}. We found a best-fit with a $\chi^2_{red}=0.91$ using the same data points as \citet{torus+jet}. The values are reported in Table~\ref{tab:quiescent table params} and Fig.~\ref{fig:quiescent+spectrum} shows the associated spectrum and the image at $2.2\ \mu m$. We obtain a magnetic field strength of $257\ \mathrm{G}$ for the jet and $212\ \mathrm{G}$ at the center of the torus.
These values are higher than in \citet{2019ApJ...881L...2B} who considers a full thermal electron population with a higher temperature but are of the same order as in \citet{2022MNRAS.511.3536S}.

\section{Ray-tracing setup}\label{sec:GYOTO}
We consider a Kerr black hole with dimensionless spin parameter $a=0$, described in Boyer-Lindquist ($t, r, \theta, \varphi$) coordinates. We work in units where the gravitational constant and the speed of light are equal to 1, $G = c = 1$. Radii are thus expressed in units of the black hole mass $M$.

We use the backward ray-tracing code \texttt{GYOTO}\footnote{\href{https://gyoto.obspm.fr}{https://gyoto.obspm.fr}} \citep{GYOTO, paumard_thibaut_2019_2547541} to compute images of our models at different epochs. Each pixel of our image corresponds to a direction on sky. For each pixel of the image (i.e. each direction), a null geodesic is integrated backwards in time from the observer towards the black hole, integrating along this path the radiative transfer equation

\begin{equation}
    \frac{d I_\nu}{ds} = -\alpha_\nu I_\nu + j_\nu
    \label{eq:radiative transfer}
\end{equation}

using the synchrotron emissivity $j_\nu$ and absorptivity $\alpha_\nu$ coefficients, considering various electron distribution functions. This allows us to determine the flux centroid for each epoch and trace its motion. In addition to astrometry we also determine the total flux emitted as the sum of the intensity weighted by the element of solid angle subtended by each pixel, again, for each epoch which allows us to plot the light curve.

The images produced are 1000x1000 pixels with a field of view of $300\ \mu as$ vertically and horizontally which makes a resolution $< 0.1\ \mu as^2$/pixel. This high resolution is needed to resolve properly the secondary image which has a very important role in both astrometry and light curve (see Sect.~\ref{sec:hot spot+ jet}).

We model the quiescent state of Sgr A* at $2.2\ \mu m$ with a jet (see Sect.~\ref{sec:quiescent model}). However, computing an image of the jet is $\sim 200$ times longer than an image of the flare source (i.e. the hot spot or the plasmoid, Sect.~\ref{sec:hot spot+ jet} and \ref{sec:Plasmoid} respectively) because the jet is much more extended, and integrating the radiative transfer equation is thus much longer. The absorption in the jet is negligible thus the flux emitted by the flare which crosses the jet is fully transmitted. We can compute a single image of the jet that we add to each images of the hot spot a posteriori. We then calculate the total flux by summing the jet flux with the one of the flare at a given time. The final centroid position is calculated by a simple barycenter of the two centroids (jet and flare).

\section{Intrinsic emission of the Plasmoid} \label{ap:intrinsic emission}
\subsection{Tests on the kinetic simulations}\label{sec:Plasmoid_test}
In our model, we follow the evolution of the electron distribution taking into account the injection of accelerated electrons by the merging of small plasmoids into our large plasmoid and their cooling through synchrotron radiation. The emissivity $j_\nu$ and absorptivity $\alpha_\nu$ coefficients, needed to integrate the radiative transfer Eq.~\ref{eq:radiative transfer}, are computed through the formula of \citet{Chiaberge, Rybicki} (with our notation)
\begin{equation}
    j_\nu(t) = \frac{1}{4\pi} \int_{\gamma_{min}}^{\gamma_{max}} d\gamma N_e(\gamma,t) P_s(\nu,\gamma),
    \label{eq:j_nu}
\end{equation}
\begin{equation}
    \alpha_\nu(t)=-\frac{1}{8 \pi m_e \nu^2} \int_{\gamma_{min}}^{\gamma_{max}} \frac{N_e(\gamma,t)}{\gamma l} \frac{d}{d\gamma} [\gamma l P_s(\nu,\gamma)]
    \label{eq:a_nu}
\end{equation}
with 
\begin{multline}
    P_s(\nu,\gamma)=\frac{3\sqrt{3}}{\pi} \frac{\sigma_T c U_B}{\nu_B}x^2 \\ 
    \left\{ K_{4/3} (x) K_{1/3} (x)-\frac{3}{5}x [K_{4/3}^2(x)-K_{1/3}^2(x)]  \right\}
    \label{eq:single_elec_emis}
\end{multline}
where $l=(\gamma^2-1)^{1/2}$ is the electron momentum in units of $m_e c$, $x=\nu / (3 \gamma^2 \nu_B)$, $\nu_B=eB/(2\pi m_e c)$ and $K_a(t)$ is the modified Bessel function of order $a$. We note that the Eq.~\ref{eq:single_elec_emis}, is already averaged over pitch angle. For standard distributions as thermal, power-law and $\kappa$-distributions, these formulae are equivalent to the fits of \citet{2016ApJ...822...34P} (see Appendix~\ref{ap:synchro}) that we used for computing the quiescent synchrotron flux.

As electrons start to cool as soon as they are injected in the plasmoid, the full distribution is no more a $\kappa$-distribution. However, turning off the cooling during the growth phase allows us to compare the results of \texttt{EMBLEM} to the fitting formulae of \citet{2016ApJ...822...34P}. As we inject electrons following their definition of the $\kappa$-distribution with a linear increase of the number density, the two approach show similar results (see Appendix~\ref{ap:synchro}). In our cases, the absorption is very low, thus we neglect the absorption in these tests. We can derive an analytical formula for the specific intensity from the high frequency limit emissivity Eq.~\ref{eq:Js,hi} depending on the number density $n_e$, the electron temperature $\Theta_e$ and the magnetic field $B$ in case the cooling is switched off during growth. We find in the case without cooling that, keeping $\kappa$ constant 
\begin{equation}
    I_{\nu,max} \propto n_{e,max} \ \Theta^{\ \kappa-2} \ B^{\ \kappa/2} \text{, if } X_\kappa > 2000.
    \label{eq:analytic_high_freq}
\end{equation}

\begin{figure}
    \resizebox{\hsize}{!}{\includegraphics{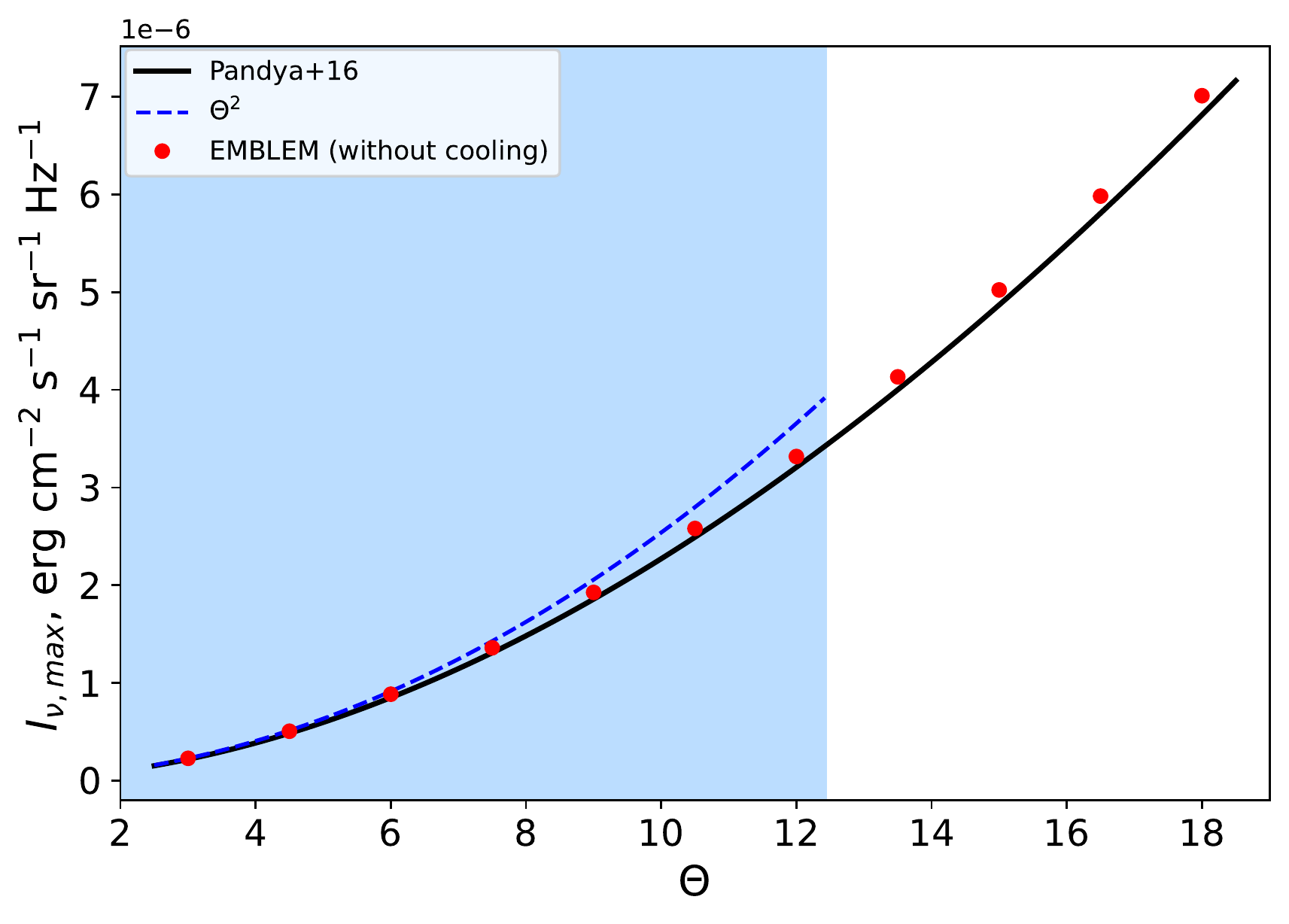}}
    \caption{Specific intensity at the end of the growth phase ($t=t_\mathrm{growth}=75$ $r_g/c$) of a $\kappa$-distribution with $n_e=5 \times 10^6$ cm$^{-3}$, $B=10$ G, $\kappa=4$ for a range of $\Theta_e$ computed from the full fitting formulae of \citet{2016ApJ...822...34P} (black curve), by the \texttt{EMBLEM} code (red dots) and with the high-frequency limit analytical expression (dashed blue curve). We overplot in light blue the range of $\Theta_e$ where $X_\kappa > 2000$ i.e. where the relative error between the high frequency limit and the full formula is lower than $20 \%$.}
    \label{fig:Comparison_Panday_EMBLEM_Theta}
\end{figure}

\begin{figure}
    \resizebox{\hsize}{!}{\includegraphics{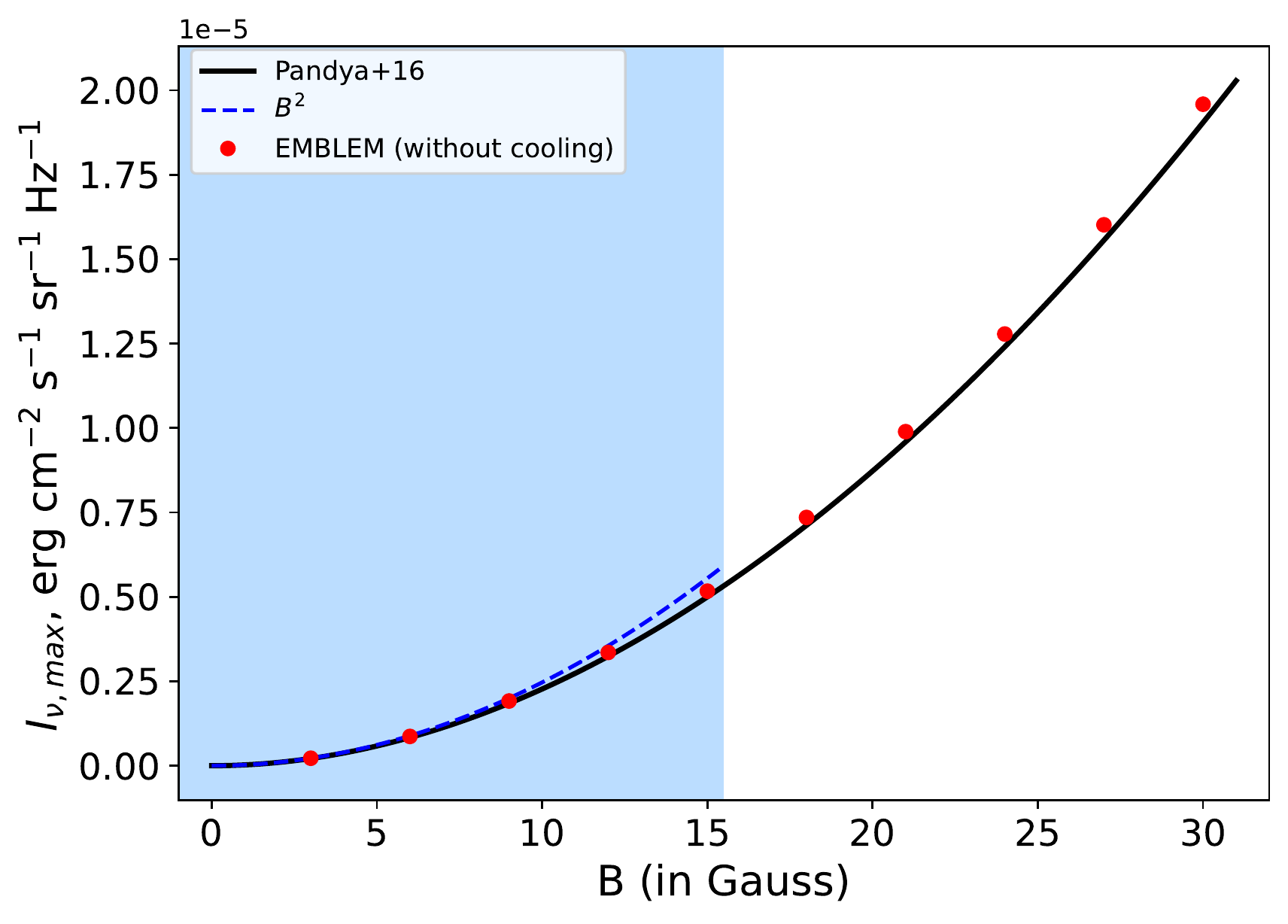}}
    \caption{Same as in Fig.\ref{fig:Comparison_Panday_EMBLEM_Theta} for a range of $B$ and with $\Theta_e=10$.}
    \label{fig:Comparison_Panday_EMBLEM_BB}
\end{figure}

We show the relative error of the maximum specific intensity between the \texttt{EMBLEM} code (red dots) and the formulae of \citet{2016ApJ...822...34P} (black curve) depending on the electron temperature and the magnetic field in Fig.~\ref{fig:Comparison_Panday_EMBLEM_Theta} and \ref{fig:Comparison_Panday_EMBLEM_BB}. We fix the others parameters to $n_e=5 \times 10^6$ cm$^{-3}$, $\kappa=4$, $t_\mathrm{growth}=75$ $r_g/c$. The values of \texttt{EMBLEM} are in good agreement with the previous analytical expression (Eq.~\ref{eq:analytic_high_freq}) for low values of $\Theta_e$ and $B$. For high values, we are beyond the validity of our approximations (in the intermediate frequency regime of the fitting formula, see Appendix~\ref{ap:synchro}). Comparing the results of \texttt{EMBLEM} (without cooling) with the full fitting formula of \citet{2016ApJ...822...34P} (black curves) results in an error lower than $5\%$ showing the good agreement between the two approaches.

\subsection{Analytical estimate of the intrinsic light curve} \label{sec:plasmoid_analytic}
Next, we compute the light curve emitted by the plasmoid which will be affected by the relativistic effects. To reproduce a given light curve, we can estimate the values of the parameters through characteristic scales. The growth time, which is a "free" parameter of the model, can be estimated from the light curve taking into account the beaming effect and thus, depends on the orbital parameters. The synchrotron cooling time of an electron with Lorentz factor
$\gamma$ in a magnetic field $B$ reads
\begin{equation}
    t_{cool} =\frac{3}{4} \frac{ m_e c}{\sigma_T U_B \gamma}
    \label{eq:cooling_time}
\end{equation}
with $\sigma_T$ the electron Thomson cross section and $U_B$ the magnetic energy density. In a Dirac spectrum approximation, the Lorentz factor of an electron emitting an IR photon at $2.2\ \mu m$ is \citep{1979rpa..book.....R}
\begin{equation}
    \bar{\gamma} = \left( \dfrac{\nu m_e c}{\eta e B} \right)^{1/2}
\end{equation}
with $\eta = (0.29 \times 3)/(2\pi)$ a dimensionless numerical factor (see Appendix~\ref{ap:sed}). One can thus constrain the magnetic field from the synchrotron cooling time as 
\begin{equation}
    t_{cool} = 19 \times \left(\frac{B}{20 G} \right)^{-1.5}\ \left(\frac{\lambda}{2.2 \mu m} \right)^{0.5} \text{min.}
\end{equation}

Taking into account the cooling of the electrons during the growth phase leads to a lower flux than what we estimate from Eq.~\ref{eq:analytic_high_freq}. Indeed, as electrons start to cool directly after being injected, the integral of the distribution in Eq.~\ref{eq:j_nu} and so the emissivity will always be lower than without cooling. The key parameter of synchrotron cooling is the cooling time scale (Eq.~\ref{eq:cooling_time}), which depends on the magnetic field strength and the initial energy of the electrons. It has to be compared to the growth time. Indeed, with a low growth time, only high-energy electrons have the time to cool. Increasing the growth time will allow lower energy electrons to cool and so decrease even more the maximum flux of the light curve.

With some approximations (see Appendix~\ref{ap:analytlcmaximum} for the details), one can estimate the flux with cooling at $t=t_{\mathrm{growth}}$

\begin{equation} \label{eq:synSEDflareapp}
    \nu F_{\nu}^{\mathrm{syn}}(\nu,t) = \dfrac{n_e R_b^3 \, \bar{\gamma} m_e c^2}{12 D^2 t_{\mathrm{growth}} \, \kappa \theta^2}
    \begin{cases}
      \left[ \Psi(\bar{\gamma}) - \Psi(\xi(\bar{\gamma},t)) \right], & \text{for } \nu < \tilde{\nu}(t) \\
        \Psi(\bar{\gamma}), & \text{for } \nu \geq \tilde{\nu}(t)
    \end{cases}
\end{equation} 
  
where $\tilde{\nu}(t) = (\eta e B)/(m_e c b_c^2 t^2)$ is the frequency corresponding to the condition $\bar{\gamma} = 1/(b_c t)$ and
\begin{equation} \label{eq:psi_function}
\Psi(x) = \left(1 + \dfrac{x - 1}{\kappa \theta} \right)^{-\kappa} \, \left[x^2 (\kappa - 1) + 2x (\kappa\theta - 1) + 2\theta (\kappa \theta - 2) \right].
\end{equation}

\begin{figure}
    \resizebox{\hsize}{!}{\includegraphics{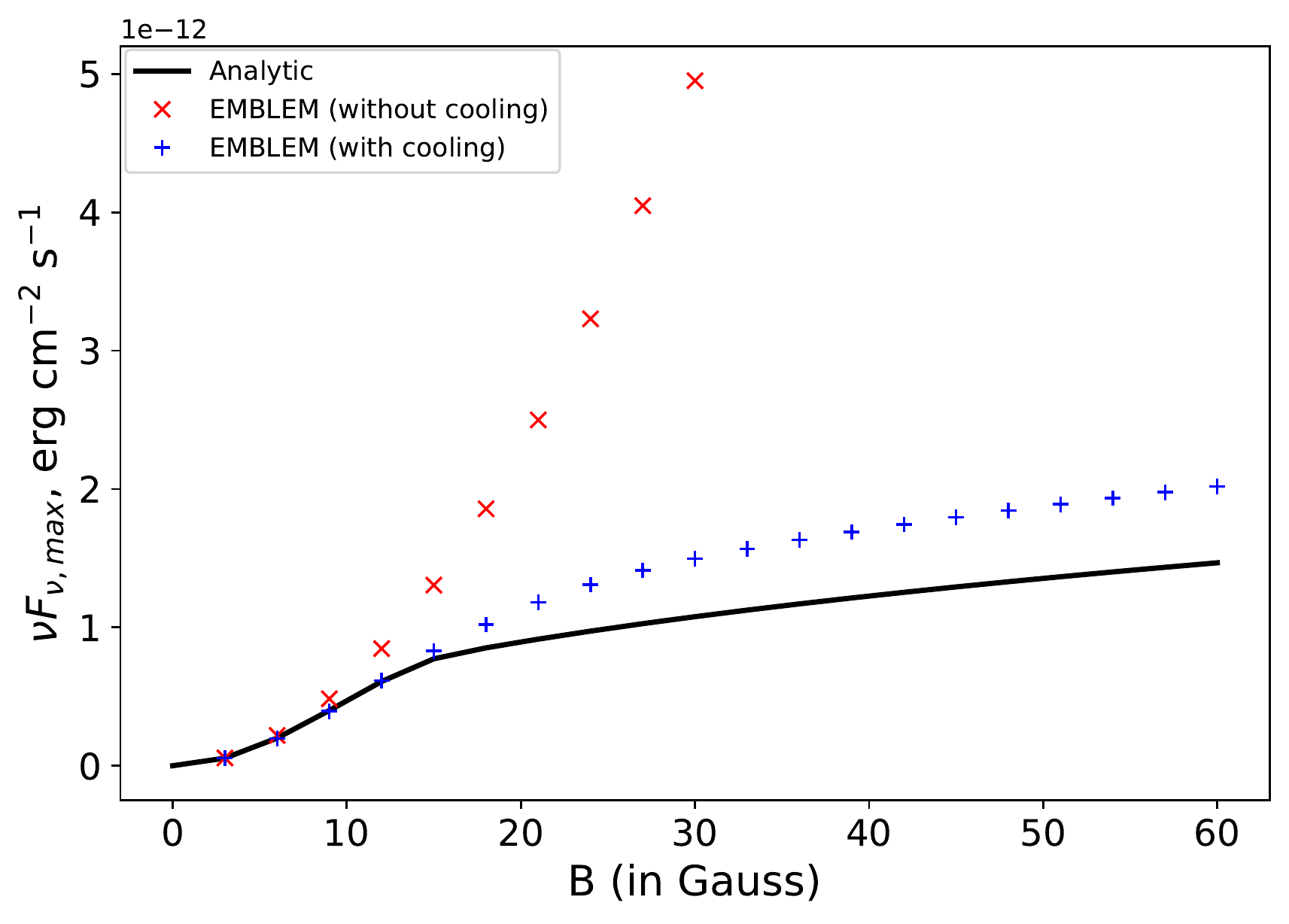}}
    \caption{Evolution of the maximum flux $\nu F_\nu (t_\mathrm{growth})$ (at the end of the growth phase $t_\mathrm{growth}=75$ $r_g/c$) as a function of the magnetic field. We show the results of EMBLEM without cooling (red crosses) as in Fig.~\ref{fig:Comparison_Panday_EMBLEM_BB}. Allowing the cooling during the growth phase results in a lower maximum flux (blue crosses). One can estimates the maximum flux with cooling through Eq.~\ref{eq:synSEDflareapp} (see Appendix~\ref{ap:analytlcmaximum} for details). This equation is divided in two regimes, the equilibrium regime where the magnetic field is strong enough to compensate the injection and creates a stationary state ($B \geq 16.2$ G) and non stationary regime where not all electrons has cooled at $t_\mathrm{growth}$ ($B< 16.2$ G). The relative error between the analytical formula and the results of EMBLEM (with cooling) is below 30$\%$ in the whole domain and below 7$\%$ in the non stationary regime.}
    \label{fig:EMBLEM_cooling}
\end{figure}

We plot the maximum light curve evolution relative to the magnetic field with \texttt{EMBLEM} with (blue crosses) and without (red crosses) cooling during the growth phase and the previous analytical expression in Fig.~\ref{fig:EMBLEM_cooling} (black line). As expected, the cooling becomes more significant with a strong magnetic field until the maximum flux starts to decrease for very high values ($B > 100$ G). The two regimes of Eq.~\ref{eq:synSEDflareapp} are clearly visible in Fig.~\ref{fig:EMBLEM_cooling} with a turning point at $B=16.2$ G. This approximation has a maximum relative error lower than $30\%$ compared to the results of \texttt{EMBLEM} in the stationary regime and below $7\%$ for the non stationary regime making it a good approximation estimate the peak light curve flux.

\section{Computation of the synchrotron coefficients for the Plasmoid}\label{ap:synchro}
\subsection{Fitting formulae of \citet{2016ApJ...822...34P}}
In the hot spot model and for the test of \texttt{EMBLEM}, we used the fitting formula of \citet{2016ApJ...822...34P} to compute the emissivity $j_\nu$ and absorptivity $\alpha_\nu$ considering a well defined $\kappa$-distribution. This distribution has two regimes, the low and high frequency regimes.

In the low frequency limit, the emissivity is
\begin{equation}
    j_{\nu,low} = \frac{n_e e^2 \nu_B}{c}\ X_\kappa^{1/3}\ sin(\theta)\ \frac{4 \pi \Gamma (\kappa-4/3)}{3^{7/3} \Gamma(\kappa-2)}\\
    \label{eq:Js,lo}
\end{equation}
and the absorption coefficient is
\begin{multline}
    \alpha_{\nu,low} = \frac{n_e e^2}{\nu m_e c}\ X_\kappa^{-2/3} 3^{1/6} \frac{10}{41}\ \frac{2 \pi}{(\Theta_e \ \kappa)^{10/3-\kappa}} \frac{(\kappa-2)(\kappa-1)\kappa}{3\kappa-1} \\
    \times \Gamma \left( \frac{5}{3} \right) _2F_1 \left( \kappa-\frac{1}{3}, \kappa+1,\kappa+\frac{2}{3}, -\Theta_e \ \kappa \right)
    \label{eq:As,lo}
\end{multline}
where $_2F_1$ is the hypergeometric function.

In the high-frequency limit, the emissivity is
\begin{multline}
    j_{\nu,high} = \frac{n_e e^2 \nu_B}{c}\ X_\kappa^{-(\kappa-2)/2}\ sin(\theta)\ 3^{(\kappa-1)/2} \\
    \times \frac{(\kappa-2)(\kappa-1)}{4}\ \Gamma \left( \frac{\kappa}{4} - \frac{1}{3} \right) \Gamma \left( \frac{\kappa}{4} + \frac{4}{3} \right)
    \label{eq:Js,hi}
\end{multline}
and the absorption coefficient is
\begin{multline}
    \alpha_{\nu,high} = \frac{n_e e^2}{\nu m_e c}\ X_\kappa^{-(1+\kappa)/2}\ \frac{\pi^{3/2}}{3}\  \frac{(\kappa-2)(\kappa-1)\kappa}{(\Theta_e \ \kappa)^3} \\
    \times \left( \frac{2 \Gamma(2+\kappa/2)}{2+\kappa} - 1 \right)\ \left( \left(\frac{3}{\kappa}\right)^{19/4} +\frac{3}{5} \right).
    \label{eq:As,hi}
\end{multline}

The final approximations for the emissivity and absorption coefficient are
\begin{equation}
    j_\nu = \left( j_{\nu,low}^{-x_j} + j_{\nu,high}^{-x_j} \right)^{-1/x_j}
\end{equation}
\begin{equation}
    \alpha_\nu = \left( \alpha_{\nu,low}^{-x_\alpha} + \alpha_{\nu,high}^{-x_\alpha} \right)^{-1/x_\alpha}
\end{equation}
with $x_j=3\kappa^{-3/2}$ and $x_\alpha=\left( -\frac{7}{4} + \frac{8}{5}\kappa \right)^{-43/50}$.

\begin{figure}
    \resizebox{\hsize}{!}{\includegraphics{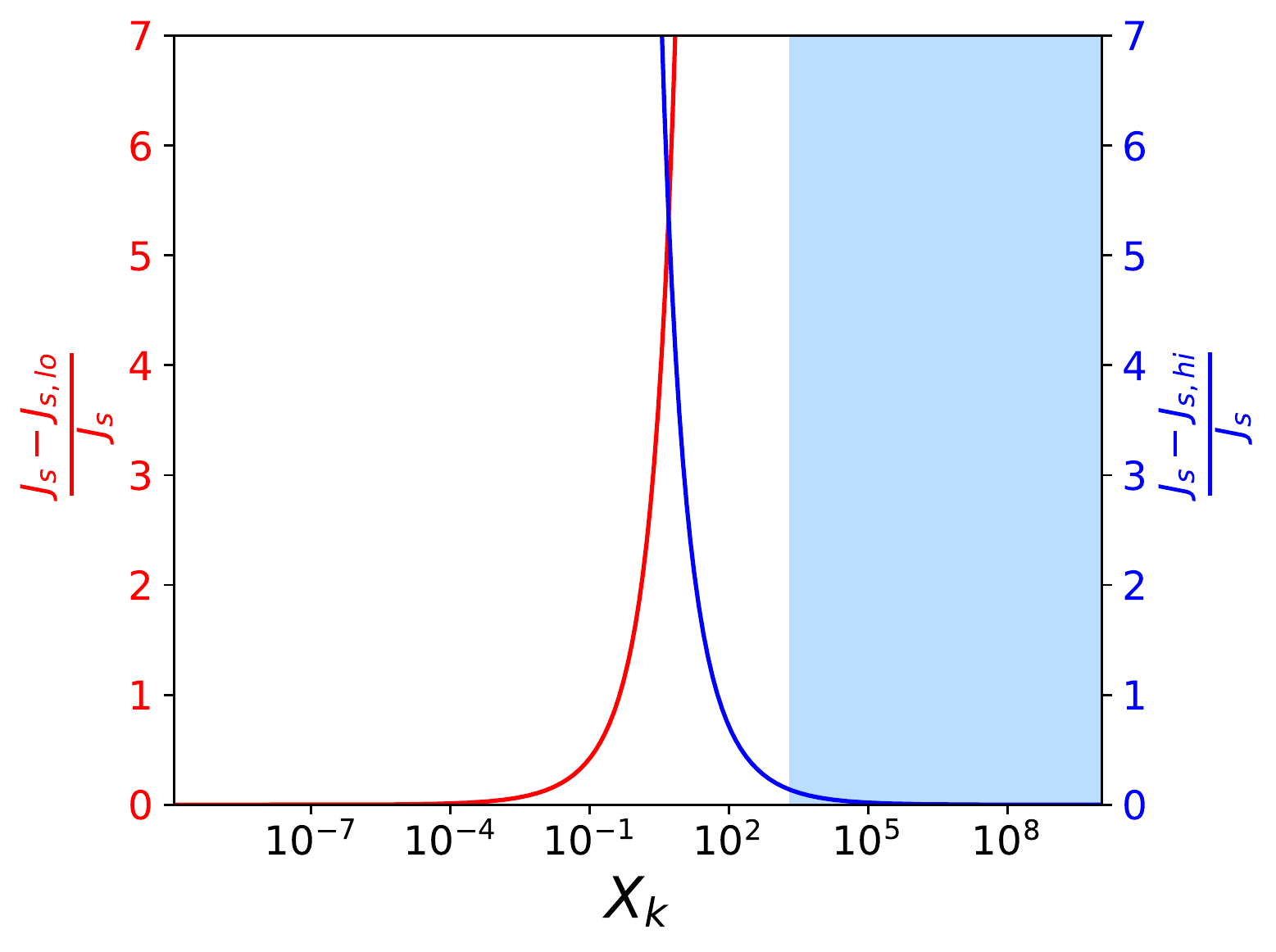}}
    \caption{Relative error between the low frequency regime (in red) - resp. the high frequency regime (in blue) - fit formulae $J_{s,lo}$ (resp. $J_{s,hi}$) of \citet{2016ApJ...822...34P} and the full fit formula of the emission coefficient $J_s$ in function of $X_\kappa = \frac{\nu}{\nu_\kappa}$ with $\nu_\kappa=\nu_B \ (\Theta_e \kappa)^2$.}
    \label{fig:Xk_error}
\end{figure}

The two frequency limits do not have the same dependence on the parameters. The frequency regime is defined by the dimensionless parameter $X_\kappa=\nu/\nu_\kappa$, with $\nu_\kappa= \nu_B (\Theta_e \kappa)^2$. Fig.~\ref{fig:Xk_error} shows the relative error of the two regimes (the low frequency in red and the high frequency in blue) compared to the final emission coefficient. While at very high (respectively very low) $X_\kappa$, the high frequency (resp. low frequency) fitting formulae work very well, there is a large frequency regime ($10^{-2} \lesssim X_\kappa \lesssim 10^3$), hereafter \textit{intermediate regime}, where both limits are needed. At $2.2$ $\mu m$, $X_\kappa > 1$, while $\Theta_e \ \kappa \lesssim 10 ^3$ which correspond to our typical set of parameters. This is why we used the high frequency regime for our test our \texttt{EMBLEM}.

\subsection{\citet{Chiaberge} approximation}
Modeling the synchrotron cooling of the electrons with a thermal, power-law or $\kappa$ distribution is not trivial. Indeed, the evolution of the energy of an electron which emits synchrotron radiation is (e.g., \citet{Rybicki})

\begin{equation} \label{eq:elenevol}
    \gamma(t) = \gamma_0 (1+A\gamma_0t)^{-1}
\end{equation}
\begin{equation}
    \text{with } A=\frac{4}{3} \frac{\sigma_T B^2}{8 \pi m_e c},
\end{equation}
$\gamma$ the Lorentz factor of the electron at time $t$, and $\gamma_0$ the initial Lorentz factor. The energy evolution strongly depends on the initial energy. The higher the initial energy of the electron, the faster it will cool. Thus, the initial distribution we could impose will quickly be deformed (see top-left panel of Fig.~\ref{fig:ElecDistribEvol}) and could not be modeled by one (or more) of the three distribution of \citet{2016ApJ...822...34P} (thermal, power-law and/or $\kappa$).

In order to properly model the cooling of electrons, we simulate the evolution of the electron distribution with injection and synchrotron cooling (Sect.~\ref{sec:Plasmoid}). These simulations give us the electron distribution $N_e(\gamma,t)$ at different times. We compute the emissivity $j_\nu$ and the absorptivity $\alpha_\nu$ associated for a range of frequencies from $10^6$ to $10^{21}$ Hz following the formula of \citet{Chiaberge} (with our notation)
\begin{equation}
    j_\nu(t) = \frac{1}{4\pi} \int_{\gamma_{min}}^{\gamma_{max}} d\gamma N_e(\gamma,t) P_s(\nu,\gamma)
\end{equation}
and the absorption coefficient follows
\begin{equation}
    \alpha_\nu(t)=-\frac{1}{8 \pi m_e \nu^2} \int_{\gamma_{min}}^{\gamma_{max}} \frac{N_e(\gamma,t)}{\gamma p} \frac{d}{d\gamma} [\gamma p P_s(\nu,\gamma)],
\end{equation}
where $p=(\gamma^2-1)^{1/2}$ is the electron momentum in units of $m_e c$ and $P_s$ is the emissivity of a single electron (see~\ref{eq:single_elec_emis}).

In order to obtain the emissivity and absorption coefficient at any time and any frequency (to account the relativistic Doppler effect for example), we made a bilinear interpolation.

\section{Analytical approximation for Sgr A* flare peak flux} \label{ap:analytlcmaximum}

Here we derive an analytical expression to compute the time-dependent flux from Sgr A* flares during the growth phase, and obtain an analytical formula for the peak flare flux. For that, we first obtain the approximate analytical form of the varying electron spectrum during the growth phase by solving the kinetic equation, and then compute the approximate synchrotron SED associated to the time-dependent electron spectrum.

\subsection{Deriving time-dependent electron spectrum during the growth phase}

The kinetic equation describing the evolution of the electron spectrum during the growth phase is given by Eq.~\ref{eq:kineticeq}:

\begin{equation} \label{eq:kineticeqaaip}
    \dfrac{\partial N_e(\gamma,t)}{\partial t} = \dfrac{\partial}{\partial \gamma} \left(b_c \gamma^2 N_e(\gamma,t) \right) \, + \, Q_{\mathrm{inj}}(\gamma,n_e,\theta,\kappa)
\end{equation}

with the injection term $Q_{\mathrm{inj}}(\gamma)$ given by Eq.~\ref{eq:injterm} and Eq.~\ref{eq:kappadistribut}, and synchrotron cooling term $\dot{\gamma}_{\mathrm{syn}} = -b_c (\gamma^2 - 1)$ (see Eq.~\ref{eq:kinetic}), where $b_c = (4 \sigma_T U_B)/(3 m_e c)$. We use here an approximation $\dot{\gamma}_{\mathrm{syn}} \approx -b_c \gamma^2$, as the bulk of the electrons producing the flare emission are highly relativistic.

We use the method of characteristics to solve the kinetic equation. We search for characteristic curves in the $\gamma$-t space, along which our differential equation in partial derivatives becomes an ordinary differential equation. Let us rewrite the kinetic equation in the following form, expanding the derivative on the Lorentz factor:

\begin{equation} \label{eq:kineticrewritten}
    \dfrac{\partial N_e(\gamma,t)}{\partial t} \, + \, (-1) b_c \gamma^2 \, \dfrac{\partial N_e(\gamma,t)}{\partial \gamma}  = Q_{\mathrm{inj}}(\gamma) + 2 \gamma b_c N_e(\gamma,t)
\end{equation}

When restricting our equation to the characteristic curve ($\gamma(t)$,$t$), the full derivative of the electron spectrum over time, by the chain rule, is:

\begin{equation}
    \dfrac{dN_e(\gamma,t)}{dt} \, = \, \dfrac{\partial N_e(\gamma,t)}{\partial t} \, + \, \dfrac{d\gamma}{dt} \, \dfrac{\partial N_e(\gamma,t)}{\partial \gamma}
\end{equation}

Comparing this to Eq.~\ref{eq:kineticrewritten}, we identify $(-1) b_c \gamma^2 = \dfrac{d\gamma}{dt}$, and therefore along the chosen characteristic curve, our equation is split into a system of two ordinary differential equations:

\begin{equation}
    \begin{cases}
     d\gamma/dt = - b_c \gamma^2\\
     dN_e(\gamma,t)/dt = Q_{\mathrm{inj}}(\gamma) + 2 \gamma b_c N_e(\gamma,t)
    \end{cases}
\end{equation}

\vspace{3mm}

The solution of the first equation is (applying the initial condition that $\gamma(t=0) = \xi$):

\begin{equation} \label{eq:gammavstchc}
    \gamma(t) = \dfrac{1}{b_c t \, + \, 1/\xi}
\end{equation}

This equation defines a characteristic curve in the $\gamma$-t space. We have chosen the initial point of the characteristic curve as $(\xi,0)$. The physical meaning of $\xi$ is the initial value of the Lorentz factor of an electron before it starts undergoing the cooling process. Eq.~\ref{eq:gammavstchc} is equivalent to Eq.~\ref{eq:elenevol}, and describes how the Lorentz factor of an individual electron evolves in time due to synchrotron cooling. From this equation, the initial Lorentz factor $\xi$ is:

\begin{equation} \label{eq:xiinitlf}
    \xi = \xi(\gamma,t) = \dfrac{1}{1/\gamma - b_c t}
\end{equation}

This formula defines the initial Lorentz factor of the characteristic curve that passes through a point ($\gamma$,t). We denote the function $N_e(\gamma_{\xi}(t),t) = u(t)$ (electron spectrum along the characteristic curve), and solve the second equation in the system:

\begin{equation}
    du/dt - 2 b_c \gamma(t) u = Q_{\mathrm{inj}}(\gamma(t))
\end{equation}

The generic solution of this linear differential equation is:

\begin{equation}
    u(t) = \dfrac{1}{\mu(t)} \, \int_{0}^{t} \, \mu(t^{\prime}) \, Q_{\mathrm{inj}}(\gamma(t^{\prime})) \, dt^{\prime} \, + \, \dfrac{C}{\mu(t)}
\end{equation}

with $C$ being the integration constant, and the function $\mu(t)$ being the integration factor, which is equal to:

\begin{equation}
   \mu(t) = \mathrm{exp}\left(\int -2 b_c \gamma(t) dt \right) \, = \, \dfrac{1}{(b_c t + 1/\xi)^2}
\end{equation}

As the electron spectrum at $t=0$ is zero, we set the initial condition $u(t=0)=0$, which results in $C=0$. Therefore, the solution for $u(t)$ is:

\begin{equation}
    u(t) \, = \, (b_c t + 1/\xi)^2 \, \int_{0}^{t} \, (b_c t^{\prime} + 1/\xi)^{-2} \, Q_{\mathrm{inj}}(\gamma(t^{\prime})) \, dt^{\prime}
\end{equation}

Now we have to return back from $u(t)$ to $N_e(\gamma,t)$, which is achieved by substitution of the equation for $\xi = \xi(\gamma,t)$ (Eq.~\ref{eq:xiinitlf}) to the expression for $u(t)$. After doing that, we obtain an expression for the electron spectrum at a moment of time $t$:

\begin{equation} \label{eq:esvstinteg}
    N_e(\gamma,t) = \dfrac{1}{\gamma^2} \int_{0}^{t} \Gamma^2 \, Q_{\mathrm{inj}}(\Gamma) \, dt^{\prime}
\end{equation}

with $\Gamma = \Gamma(\gamma,t,t^{\prime}) = \left[1/\gamma + b_c (t^{\prime} - t)\right]^{-1}$. We use here an approximation for $Q_{\mathrm{inj}}(\Gamma)$, and more specifically, for the kappa distribution, to enable analytical integration. As we are in the relativistic regime, and the peak of the injection spectrum in our case typically occurs at Lorentz factors $\gamma \gg 1$, we can substitute $\gamma (\gamma^2 - 1)^{1/2}$ with $\gamma^2$ in the Eq.~\ref{eq:kappadistribut}. This leads to some inaccuracies only at very low Lorentz factors, which virtually do not contribute to the integral value, and do not contribute to the light curve flux. We therefore use for the injected spectrum:

\begin{equation}
    Q_{\mathrm{inj}}(\gamma,n_e,\theta,\kappa) \approx \frac{N}{t_{\mathrm{growth}}} \gamma^2 \left(1 + \frac{\gamma - 1}{\kappa \theta} \right)^{-(\kappa+1)} 
\end{equation}

Now we can perform the analytical integration. We use the variable substitution from $t^{\prime}$ to $\Gamma(\gamma,t,t^{\prime})$. In this case, the differential $dt^{\prime} = - b_c^{-1} \Gamma^{-2} d\Gamma$. Our integral (Eq.~\ref{eq:esvstinteg}) then becomes:

\begin{multline}
    N_e(\gamma,t) = \dfrac{N}{\gamma^2 t_{\mathrm{growth}}} \int_{0}^{t} \Gamma^4 \left(1 + \dfrac{\Gamma - 1}{\kappa \theta} \right)^{-(\kappa+1)} \, dt^{\prime} = \\ = - \dfrac{N}{b_c \gamma^2 t_{\mathrm{growth}}} \, \int_{0}^{t} \Gamma^2 \left(1 + \dfrac{\Gamma - 1}{\kappa \theta} \right)^{-(\kappa+1)} \, d\Gamma
\end{multline}

To solve the integral, we perform integration by parts, and we obtain:

\begin{equation}
    \int_{0}^{t} \Gamma^2 \left(1 + \dfrac{\Gamma - 1}{\kappa \theta} \right)^{-(\kappa+1)} \, d\Gamma  = - \dfrac{\theta \kappa}{(\kappa-2)(\kappa-1)} \Psi(\Gamma) \, \bigg|_{0}^{t}
\end{equation}

with 

\begin{equation} \label{eq:psifunction}
\Psi(x) = \left(1 + \dfrac{x - 1}{\kappa \theta} \right)^{-\kappa} \, \left[x^2 (\kappa - 1) + 2x (\kappa\theta - 1) + 2\theta (\kappa \theta - 2) \right]
\end{equation}

We substitute the variable back from $\Gamma$ to $t^{\prime}$, with $\Gamma(t^{\prime}=0) = (1/\gamma - b_c t)^{-1} = \xi(\gamma,t)$ and $\Gamma(t^{\prime}=t) = \gamma$, as well as substitute the expression for the injection spectrum normalization, $N = (1/2) n_e (\kappa - 2)(\kappa - 1) \kappa^{-2} \theta^{-3}$ (see Eq.~\ref{eq:kappadistribut}), and obtain:

\begin{equation}
   N_e(\gamma,t) \, = \, \dfrac{n_e}{2 \kappa \theta^2 b_c \gamma^2 t_{\mathrm{growth}}} \left[ \Psi(\gamma) - \Psi(\xi(\gamma,t)) \right]
\end{equation}

One has to consider separately a special case when the $b_c t \geq 1/\gamma$, as this leads to either $\xi \rightarrow \infty$ or $\xi < 0$. Obviously, the latter situation is non-physical, as the Lorentz factor cannot be less than unity. Qualitatively, $b_c t \geq 1/\gamma \, \rightarrow \, t \geq 1/(b_c \gamma)$ means that the evolution time of an electron is longer than its cooling time-scale, and in this regime the equilibrium between the injection and cooling is already reached. Therefore, one can easily see that the time-dependent electron spectrum in the Lorentz factor domain $\gamma \geq 1/(b_c t)$ will be ``frozen'' at the steady-state one. A steady-state solution corresponds to $\xi \rightarrow \infty$, which results in $\Psi(\xi) \rightarrow 0$ (in case $\kappa > 2$). Therefore, the final solution for the time-dependent electron spectrum during the growth phase, is:
   
\begin{equation}  \label{eq:elspecapprox}
 N_e(\gamma,t) \,  = \dfrac{n_e}{2 \kappa \theta^2 b_c \gamma^2 t_{\mathrm{growth}}}
    \begin{cases}
      \left[ \Psi(\gamma) - \Psi(\xi(\gamma,t)) \right], & \text{for } \gamma < (b_c t)^{-1} \\
      \Psi(\gamma), & \text{for } \gamma \geq (b_c t)^{-1}
    \end{cases}
\end{equation}

It is worth to note, that one can obtain the same steady-state solution (the case $\gamma \geq 1/(b_c t)$) by directly solving the kinetic equation (Eq.~\ref{eq:kineticeqaaip}) with $\frac{\partial N_e}{\partial t} = 0$. To find the electron spectrum at the peak of the flare, i.e.\ at the moment when the injection is stopped, one simply calculates $N_e(\gamma,t=t_{\mathrm{growth}})$.

\subsection{Deriving time-dependent synchrotron SED during the growth phase}\label{ap:sed}

Now let us proceed to the SED and light curve computation. We use the so-called $\delta$-approximation for the electron synchrotron emissivity coefficient. This approximation assumes that a single electron with a Lorentz factor $\gamma$ emits at a single frequency, rather than a broad spectrum \citep{1979rpa..book.....R}:

\begin{equation}
    \omega_{\mathrm{peak}} \simeq 0.29 \omega_c
\end{equation}

with $\omega_c = 3 \gamma^2 eB/(m_e c)$ (averaged over pitch angles), $e$ being the electron charge, and $B$ being the magnetic field (CGS units). From this expression, one obtains:

\begin{equation} \label{eq:lambdasyndelta}
    \nu_{\mathrm{peak}} = \dfrac{\eta e \gamma^2 B}{m_e c}
\end{equation}

where $\eta = (0.29 \times 3)/(2\pi) \approx 0.14$ is a dimensionless numerical factor. For a distribution of electrons, the synchrotron SED in $\delta$-approximation, is given by \citep{dermerschl2002}:

\begin{equation} \label{eq:synseddeltaap}
    \nu F_{\nu}^{\mathrm{syn}}(\lambda) = \dfrac{4}{3} \pi R_b^3 \dfrac{c \sigma_T U_B}{6 \pi D^2} \bar{\gamma}^3 N_e(\bar{\gamma})
\end{equation}

where $R_b$ is the radius of the emitting region, $D$ is the distance between the observer and the source, and $\bar{\gamma}$ is the Lorentz factor of electrons emitting synchrotron photons with the frequency $\nu$. We obtain this Lorentz factor by expressing it from Eq.~\ref{eq:lambdasyndelta}:

\begin{equation}
    \bar{\gamma} = \left( \dfrac{m_e c \nu}{\eta e B} \right)^{1/2}
\end{equation}

Substituting the expression for $N_e(\gamma,t)$ (Eq.~\ref{eq:elspecapprox}), and the expression $b_c = (4 \sigma_T U_B)/(3 m_e c)$, into the Eq.~\ref{eq:synseddeltaap}, we finally obtain the time-dependent SED during the growth phase in $\delta$-approximation:

\begin{equation} \label{eq:synsedflareapp}
    \nu F_{\nu}^{\mathrm{syn}}(\nu,t) = \dfrac{n_e R_b^3 \, \bar{\gamma} m_e c^2}{12 D^2 t_{\mathrm{growth}} \, \kappa \theta^2}
    \begin{cases}
      \left[ \Psi(\bar{\gamma}) - \Psi(\xi(\bar{\gamma},t)) \right], & \text{for } \nu < \tilde{\nu}(t) \\
        \Psi(\bar{\gamma}), & \text{for } \nu \geq \tilde{\nu}(t)
    \end{cases}
\end{equation} 
  
where $\tilde{\nu}(t) = (\eta e B)/(m_e c b_c^2 t^2)$ is the frequency corresponding to the condition $\bar{\gamma} = 1/(b_c t)$.

\begin{figure*}[h]
    \centering
    \includegraphics[width=17cm]{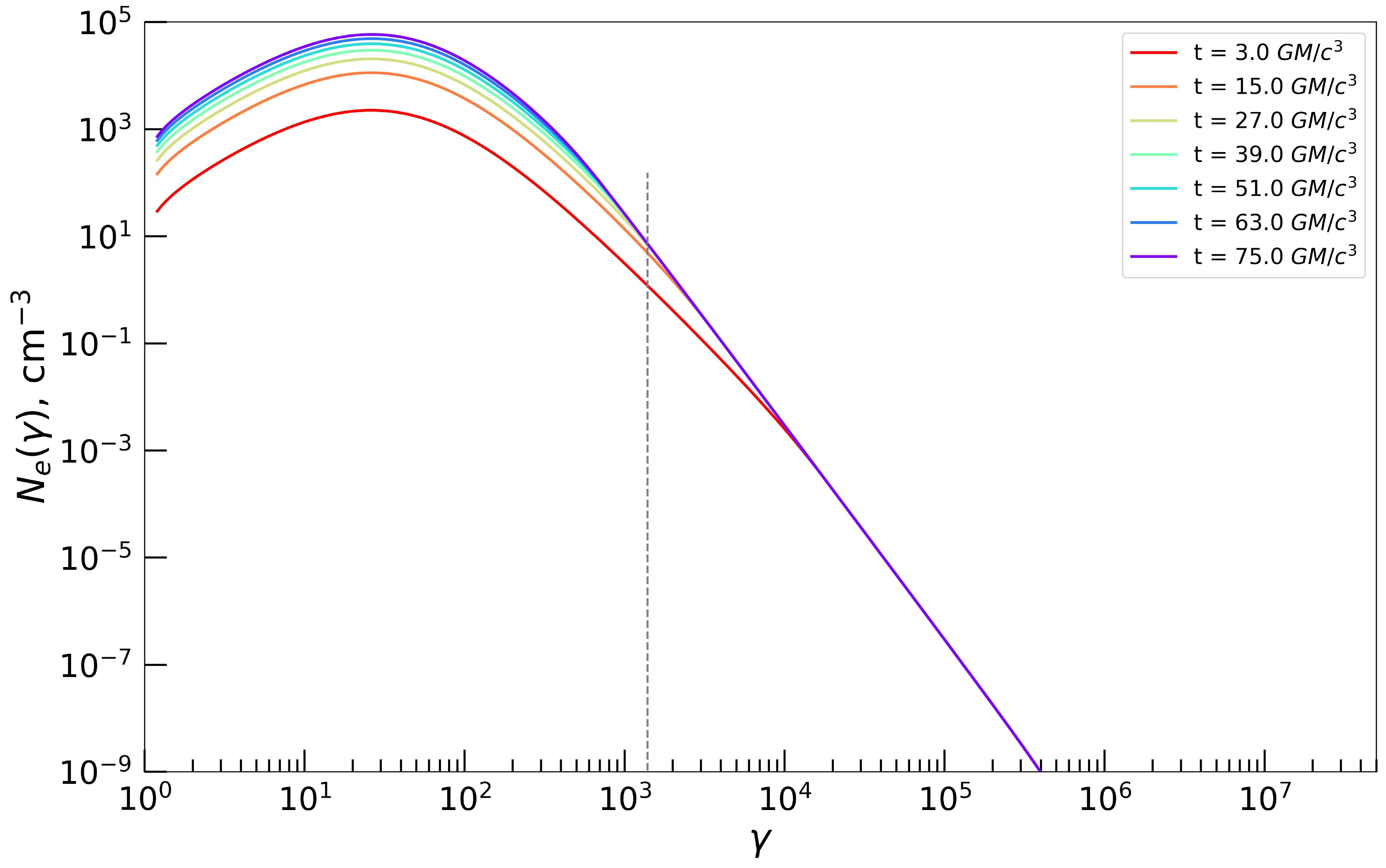}
    \caption{Time evolution of the electron distribution with \texttt{EMBLEM} (full lines) from $t=0$ to $t=t_\mathrm{growth}=75$ $r_g/c$ injecting a $\kappa$-distribution with $\Theta_e=10$ and $\kappa=4$ and $\dot{n_e}=5.10^6/t_\mathrm{growth}$. The magnetic field strength is set to 30 Gauss resulting in a stationary regime for $\gamma > 10^4$ from the very beginning. This regime extends to lower $\gamma$ values as time growths. For the estimation of the peak flux, we approximate the whole distribution (at $t=t_\mathrm{growth}$) by a simple Dirac at $\bar{\gamma}$ represented by the dashed grey line.}
    \label{fig:Distrib_approx}
\end{figure*}

\subsection{Evaluating the peak light curve flux}
To obtain a light curve during the growth phase at a specific frequency of interest $\nu_{*}$, one has to compute $\nu F_{\nu}^{\mathrm{syn}}(\nu=\nu_{*},t)$. To compute the peak light curve flux, one evaluates the quantity $\nu F_{\nu}^{\mathrm{syn}}(\nu=\nu_{*},t=t_{\mathrm{growth}})$.

\section{Additional Setup for July 22 flare} \label{ap:second_setup}
We also find another setup which reproduce well the July 22 flare data. In such scenario, the magnetic reconnection and so the plasmoid growth phase occurs way before the observing time and the flare is due to the beaming effect combined to the slow decrease of the cooling phase. The peak due to the growth phase occurs during the negative beaming part of the orbit resulting in a low flux comparable to the quiescent state.

\begin{table}[h]
    \centering
    \begin{tabular}{lcc}
        \hline
        \hline
        Parameter & Symbol & July 22 bis \\
        \hline
        \textbf{Plasmoid} & &\\
        time in \texttt{EMBLEM} at zero observing time [min] & $t_{obs,0}^{emblem}$ & $-53$\\
        initial orbital radius [$GM / c^2$] & $r_0$ & $15$ \\
        polar angle [$\degree$] & $\theta$ & $135$\\
        initial azimuthal angle [$\degree$] & $\varphi_0$ & $240$\\
        initial radial velocity [$c$] & $v_{r,0}$ & $0.01$\\
        initial azimuthal velocity [$c$] & $v_{\varphi,0}$ & $0.5$\\
        X position of Sgr A* [$\mu as$] & $x_0$ & $0$\\
        Y position of Sgr A* [$\mu as$] & $y_0$ & $0$\\
        PALN [$\degree$] & $\Omega$ & $160$\\
        \hline
    \end{tabular}
    \caption{Second orbital parameters of the plasmoid model following a conical motion used for the comparison of the July 22 flares observed by \citet{Gravity2018}.}
    \label{tab:alternativeJuly22}
\end{table}

\begin{figure*}[h]
    \centering
    \includegraphics[width=17cm]{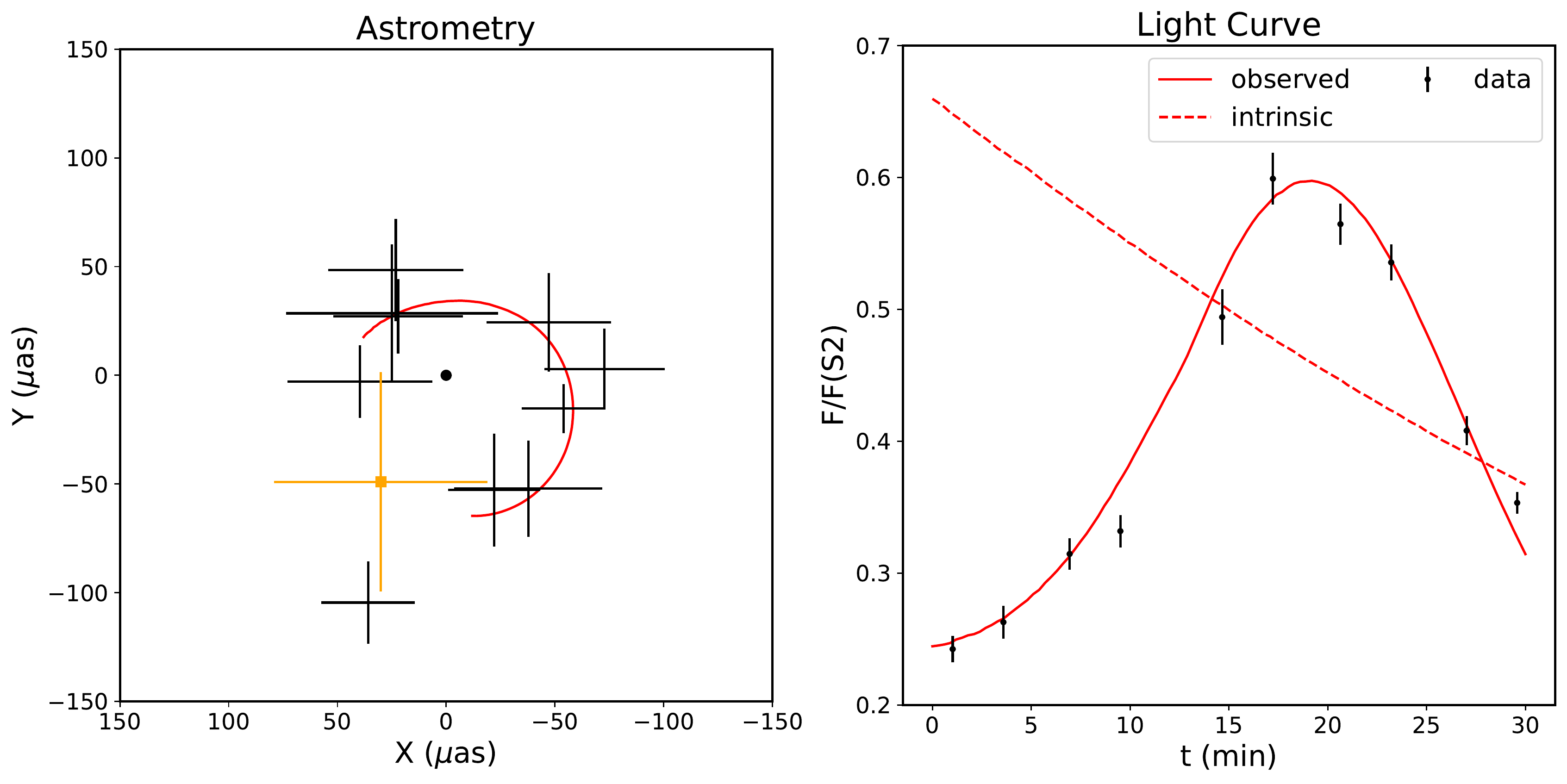}
    \caption{Data and plasmoid models of the flares from July 22, 2018. The left panels shows the astrometry of the flare while the right panel shows the light curves. Note that this is not the result of a fit. Contrary to the setup for Fig.~\ref{fig:plasmoid_flare}, the growth time is shorter $t_\mathrm{growth}=50 r_g/c$ resulting into a two peak light curve with the first one occurring at $t=-22$ min but being mitigate by the negative beaming effect. The secondary peak which match the observed flare data shown here is due to the positive beaming during the cooling phase (as shown by the intrinsic light curve). The parameter set for this model is similar to the set of July 22 and is listed in Table~\ref{tab:alternativeJuly22} with the same physical parameter as in Table~\ref{tab:inputparamemblem} but with $t_\mathrm{growth}=50 r_g/c$, $\Theta_e=72$ and $B=10$ G.}
    \label{fig:plasmoid_flare_alternative}
\end{figure*}

\end{appendix}
\end{document}